\documentclass[journal]{IEEEtran}

\usepackage{cite}
\ifCLASSINFOpdf
   \usepackage[pdftex]{graphicx}
\else
\fi
\usepackage{amsmath,amssymb,amsthm}
\usepackage{steinmetz}
\usepackage{nicefrac}
\usepackage{soul}
\usepackage{multirow}
\usepackage{balance}

\usepackage{tikz}
\usepackage{pgfplots}
\usetikzlibrary{external}
\usepgfplotslibrary{units}

\newtheorem{prop}{Proposition}
\newtheorem{definition}{Definition}
\usepackage{array}
\ifCLASSOPTIONcompsoc
  \usepackage[caption=false,font=normalsize,labelfont=sf,textfont=sf]{subfig}
\else
  \usepackage[caption=false,font=footnotesize]{subfig}
\fi

\IEEEoverridecommandlockouts

\newcommand{\fixme}[2]{\ifx&#2&{\color{red}#1}\else{\color{red}FIXME\{}#1{\color{red}\}}\footnote{{\color{red}#2}}\PackageWarning{Fixme}{#1: #2}\fi}
\usepackage{booktabs}
\usepackage{color}

\definecolor{Set1-7-1}{RGB}{228,26,28}
\definecolor{Set1-7-2}{RGB}{55,126,184}
\definecolor{Set1-7-3}{RGB}{77,175,74}
\definecolor{Set1-7-4}{RGB}{152,78,163}
\definecolor{Set1-7-5}{RGB}{255,127,0}
\definecolor{Set1-7-6}{RGB}{166,86,40}
\definecolor{Set1-7-7}{RGB}{0,0,0}

\newcommand{\figurewidth}{0.92}	
\newcommand{\figureheight}{0.74}	

\begin{document}

\bstctlcite{IEEEexample:BSTcontrol}

\title{On the Error Rate of the LoRa Modulation\\ with Interference}
\author{\IEEEauthorblockN{Orion Afisiadis,~\IEEEmembership{Student Member,~IEEE}, Matthieu Cotting,\\ Andreas Burg,~\IEEEmembership{Member,~IEEE}, and Alexios Balatsoukas-Stimming,~\IEEEmembership{Member,~IEEE}}%
\thanks{Orion Afisiadis, Matthieu Cotting, and Andreas Burg are with the Telecommunication Circuits Laboratory, \'{E}cole polytechnique f\'{e}d\'{e}rale de Lausanne, Switzerland (e-mail: orion.afisiadis@epfl.ch, matthieu.cotting@alumni.epfl.ch, andreas.burg@epfl.ch). Alexios Balatsoukas-Stimming was with the Telecommunication Circuits Laboratory, \'{E}cole polytechnique f\'{e}d\'{e}rale de Lausanne, Switzerland, and is currently with the Department of Electrical Engineering, Eindhoven University of Technology, Netherlands (e-mail: a.k.balatsoukas.stimming@tue.nl).}%
}

\maketitle

\begin{abstract}
LoRa is a chirp spread-spectrum modulation developed for the Internet of Things. In this work, we examine the performance of LoRa in the presence of both additive white Gaussian noise and interference from another LoRa user. To this end, we extend an existing interference model, which assumes perfect alignment of the signal of interest and the interference, to the more realistic case where the interfering user is neither chip- nor phase-aligned with the signal of interest and we derive an expression for the error rate. We show that the existing aligned interference model overestimates the effect of interference on the error rate. Moreover, we prove two symmetries in the interfering signal and we derive low-complexity approximate formulas that can significantly reduce the complexity of computing the symbol and frame error rates compared to the complete expression. Finally, we provide numerical simulations to corroborate the theoretical analysis and to verify the accuracy of our proposed approximations.
\end{abstract}

\IEEEpeerreviewmaketitle

\section{Introduction} \label{sec:intro}
The Internet of Things (IoT) will consist of billions of connected devices that have a large number of applications, such as smart metering, logistics~\cite{Adelantado2017}, and localization and tracking~\cite{Bakkali2017}. Other potential uses of IoT devices include health monitoring~\cite{Catherwood2018} and massive sensor networks for smart farming and environmental monitoring~\cite{Jawad2017}. Since these devices will mostly be low-power and they are expected to connect wirelessly with each other (or with centralized gateways), several specialized communications protocols have been proposed for IoT applications. Some examples include Sigfox, Weightless, NB-IoT, and LoRa~\cite{Raza2017},~\cite{Goursaud2015}.

LoRa specifically is a low-rate, low-power, and high-range modulation that uses chirp spread-spectrum for its physical layer~\cite{SX127x}. LoRa supports multiple spreading factors, coding rates, and packet lengths, to support a very wide range of operating signal-to-noise ratios (SNRs). The LoRa physical layer is proprietary~\cite{Seller2016}, but reverse engineering attempts~\cite{Knight2016,Robyns2018} have led to detailed mathematical descriptions~\cite{Vangelista2017}. The effect of carrier- and sampling frequency offset on LoRa digital receivers has been modeled and analyzed in~\cite{Ghanaatian2019}. Since LoRa uses the ISM band, interference from other technologies using the same band is a potential problem. More importantly, LoRa relies on LoRaWAN for the MAC layer, which uses an ALOHA-based channel access scheme in which collisions are not explicitly avoided. These collisions lead to same-technology inter-user interference which may ultimately become the capacity-limiting factor in massive IoT scenarios~\cite{Haxhibeqiri2018}. For this reason, it is of great interest and importance to study the performance of LoRa under same-technology interference.

The authors of~\cite{Bankov2017} present a mathematical network model for LoRa that includes the capture effect, i.e., the fact that a LoRa packet can be correctly decoded even under interference from another LoRa packet. A stochastic geometry framework for modeling the performance of a single gateway LoRa network is used in~\cite{Georgiou2017}. An investigation of the latency, collision rate, and throughput for LoRaWAN under duty-cycle restrictions is performed in~\cite{Sorensen2017}. Several real-world deployments of LoRa have been tested, but in order to assess the network scalability of LoRaWAN to future network densities that are expected to be orders of magnitude larger, evaluations through network simulators need to be performed. For this reason, the works of~\cite{Abeele2017,Reynders2018} added LoRa functionality to the well-known ns-3 network simulator. A simpler Python-based network simulator for the LoRa uplink was first described in~\cite{Bor2016}, and later extended for the LoRa downlink in~\cite{Pop2017}. The impact of the downlink feedback on LoRa capacity was also studied in~\cite{Centenaro2017}. A general overview and performance evaluations of LoRaWAN can be found in~\cite{Wixted2016,Haxhibeqiri2018}.

The impact of interference coming from different technologies on the performance of the LoRa modulation has received some attention in the literature. Specifically,~\cite{Orfanidis2017} studies the co-existence of LoRa with IEEE 802.15.4g, while~\cite{Reynders2016} studies the co-existence of LoRa with ultra-narrowband technologies, such as Sigfox. The impact of interference coming from other LoRa nodes has also received some attention. Specifically, the work of~\cite{Croce2017} extended the simulator of~\cite{Bor2016} in order to study the impact of imperfect orthogonality between different LoRa spreading factors. The work of~\cite{Croce2018} also examines the effect of imperfect orthogonality by examining the signal-to-interference ratio (SIR) threshold for receiving a packet correctly for all combinations of spreading factors. The SIR thresholds are derived both by simulations and experimental results and rectify the values found in~\cite{Goursaud2015}. Interference is particularly detrimental when users with the same spreading factor collide since the spreading can no longer mitigate the interference. The authors of~\cite{Feltrin2018} perform an experimental assessment of the link-level characteristics of the LoRa system, followed by a system-level simulation to assess the capacity of a LoRaWAN network. 

Convenient approximations for the bit-error rate (BER) of the LoRa modulation when transmission takes place over additive white Gaussian noise (AWGN) and Rayleigh fading channels are given in~\cite{Elshabrawy2018}, but collisions are not considered. Finally, the work of~\cite{Elshabrawy2018b}, which is most closely related to our work, provides an approximation for the BER of the LoRa modulation under AWGN and interference from a single LoRa interferer with the same spreading factor (\emph{same-SF}). Capacity planning for LoRa with the aforementioned interference model is addressed in~\cite{Elshabrawy2019}. The work in~\cite{Elshabrawy2019} is the first work that, to the best of our knowledge, analyzes the coverage of LoRa under a unified noise and interference framework. 

\subsubsection*{Contributions}
The work of~\cite{Elshabrawy2018b} made a significant first step toward understanding the behavior of LoRa under same-SF interference. In this paper, based on our own previous work of~\cite{Afisiadis2019}, we extend the interference model of~\cite{Elshabrawy2018b} to the more general (and more realistic) case where the interference is neither chip- nor phase-aligned with the signal-of-interest. We derive an expression for the symbol error rate (SER) under this new complete interference model. Moreover, we derive an approximation for the SER under the new interference model and we show that non-integer chip duration time-misalignment in particular has a significant effect on the SER. Specifically, we show that the interference model of~\cite{Elshabrawy2018b} is pessimistic in the sense that it consistently over-estimates the actual SER. We also prove two properties of same-SF LoRa-induced interference that enable a significant reduction of the complexity of calculating both the exact and the approximated SER. Similar properties which were stated (without proof) in~\cite{Elshabrawy2018b} are shown to be special cases of our results. Finally, we derive an approximation for the frame error rate (FER), which is generally of greater practical interest for network simulators such as the ones presented in~\cite{Abeele2017,Reynders2018,Bor2016,Pop2017}.

\subsubsection*{Outline}
The remainder of this paper is organized as follows. In Section~\ref{sec:system_model}, we provide a detailed description of the LoRa modulation and demodulation processes. In Section~\ref{sec:SER_AWGN}, we derive an expression and we review existing approximations for the SER of the LoRa modulation under AWGN. In Section~\ref{sec:SER_interf}, we model the behavior of LoRa under same-SF interference with neither chip- nor phase-alignment assumptions and we derive a corresponding expression for the SER. In Section~\ref{sec:symmetries}, we explore and prove the existence of equivalent interference patterns that can be exploited to reduce the complexity of computing the SER and we derive a low-complexity approximation for the SER. In Section~\ref{sec:fer} we also derive an approximation for the FER, which is of greater practical interest than the SER in network simulators. Finally, Section~\ref{sec:results} contains numerical SER and FER results and Section~\ref{sec:conclusion} concludes this paper.

\subsubsection*{Notation} Bold lowercase letters (e.g., $\mathbf{a}$) denote vectors, while bold uppercase letters (e.g., $\mathbf{A}$) denote the discrete Fourier transform (DFT) of $\mathbf{a}$, i.e., $\mathbf{A} = \text{DFT}(\mathbf{a})$. We define $[x]_y = x \mod y$. We denote the normal and complex normal (with i.i.d. components) probability density functions (PDFs) with mean $\mu$ and variance $\sigma^2$ as $\mathcal{N}(\mu,\sigma^2)$ and $\mathcal{CN}(\mu,\sigma^2)$, respectively. Moreover, we denote the PDF and the cumulative density function (CDF) of the Rayleigh and Rice distributions by $f_{\text{Ra}}(y; \sigma)$, $f_{\text{Ri}}(y; v, \sigma)$ and $F_{\text{Ra}}(y;\sigma)$, $F_{\text{Ri}}(y;v, \sigma)$, respectively, where $\sigma$ and $v$ denote the \emph{scale} and \emph{location} parameters.

\section{The LoRa Modulation} \label{sec:system_model}
In this section, we briefly summarize the LoRa modulation and how the demodulation can be performed.

\subsection{Modulation}
LoRa is a spread-spectrum modulation that uses a bandwidth $B$ and $N = 2^\text{SF}$ chips per symbol, where SF is called the \emph{spreading factor} with $\text{SF} \in \{7, \dots, 12\}$. When considering the discrete-time baseband equivalent signal, the bandwidth $B$ is split into $N$ frequency steps. A symbol $s \in \mathcal{S}$, where $ \mathcal{S} = \left\{0,\hdots,N{-}1\right\} $, begins at frequency $(\frac{s B}{N} - \frac{B}{2})$. The frequency increases by $\frac{B}{N}$ at each chip until it reaches the Nyquist frequency $\frac{B}{2}$. When the Nyquist frequency is reached, there is a frequency fold to $-\frac{B}{2}$ at chip $n_{\text{fold}} = N-s$~\cite{Chiani2019}. The general discrete-time baseband equivalent equation of a LoRa symbol $s$ is
\begin{align}
x_s[n] & =
     \begin{cases}
       e^{j2\pi\left( \frac{1}{2N} \left(\frac{B}{f_s}\right)^2n^2 + \left(\frac{s}{N} - \frac{1}{2}\right)\left(\frac{B}{f_s}\right)n \right)}, &  n \in \mathcal{S}_1,\\
       e^{j2\pi\left( \frac{1}{2N} \left(\frac{B}{f_s}\right)^2n^2 + \left(\frac{s}{N} - \frac{3}{2}\right)\left(\frac{B}{f_s}\right)n \right)}, &  n \in \mathcal{S}_2,
     \end{cases}
\end{align}
where $\mathcal{S}_1 = \{0,..., n_{\text{fold}}-1\}$ and $\mathcal{S}_2 = \{n_{\text{fold}},...,N-1\}$. In the practically relevant case where the sampling frequency $f_s$ is equal to $B$, which we assume for the remainder of this manuscript, the discrete-time baseband equivalent description of a LoRa symbol $s$ can be simplified to
\begin{align} \label{eq:LoRa_symbol}
x_s[n] & = e^{j2\pi \left(\frac{n^2}{2N}  + \left(\frac{s}{N} - \frac{1}{2}\right)n \right)}, \;\; n \in \mathcal{S}.
\end{align}
After transmission over a time-invariant and frequency-flat wireless channel with complex-valued channel gain $h \in \mathbb{C}$, the received LoRa symbol is given by
\begin{align}
  y[n] & = hx_s[n] + z[n], \;\; n \in \mathcal{S}, \label{eq:lora_rx}
\end{align}
where $z[n] \sim \mathcal{CN}(0,\sigma^2)$ is complex additive white Gaussian noise with $\sigma^{2} = \frac{N_0}{2N}$ and $N_0$ is the single-sided noise power spectral density. We assume that $|h| = 1$ without loss of generality, so that the signal-to-noise ratio (SNR) is
\begin{align}
  \text{SNR}  & = \frac{1}{N_0}. \label{eq:SNRdef}
\end{align}

\begin{figure}
	\centering
	\includegraphics[width=0.50\textwidth]{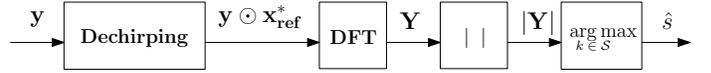}
	\caption{Illustration of a DFT-based LoRa demodulation chain.}
	\label{fig:demo_chain}
\end{figure}

\subsection{Demodulation}
To demodulate the symbols, the correlation of the received signal with all the possible symbols $k \in \mathcal{S}$ is computed as
\begin{align}
	X_k & = \sum_{n = 0}^{N-1} y[n]x_k^*[n] \\
      & = |h|\sum_{n = 0}^{N-1} e^{j 2 \pi \left( \frac{s-k}{N} \right)n + \phi} + \sum_{n = 0}^{N-1} z[n]x_k^*[n] \\
      & = |h|\sum_{n = 0}^{N-1} e^{j 2 \pi \left( \frac{s-k}{N} \right)n + \phi} + \tilde{z}_k, \label{eq:corr_demod}
\end{align}
where $\phi = \phase{h}$ denotes a phase shift introduced by the transmission channel $h$ that is fixed for each transmitted packet but generally uniformly distributed in $[0,2\pi)$, and $\tilde{z}_k \sim \mathcal{CN}(0,N\sigma^2)$. In a non-coherent receiver a symbol estimate $\hat{s}$ is obtained as
\begin{align}
  \hat{s} & = \arg\max_{k\in \mathcal{S} } \left( |X_k| \right). \label{eq:retrieved_symbol}
\end{align}
The complexity of computing~\eqref{eq:retrieved_symbol} is $O(N^2)$. The following equivalent and low-complexity method can also be used to perform the demodulation. First, a \emph{dechirping} is performed, where the received signal is multiplied by the complex conjugate of a reference signal $x_{\text{ref}}$. A convenient choice for this reference signal is an \emph{upchirp}, i.e., the LoRa symbol for $s=0$
\begin{align}
  x_{\text{ref}}[n] = e^{j2\pi \left(\frac{n^2}{2N} - \frac{n}{2} \right)}, \;\; n \in \mathcal{S}. \label{eq:Ref_symbol}
\end{align}
Then, the non-normalized discrete Fourier transform (DFT) is applied to the dechirped signal in order to obtain $\mathbf{Y} = \text{DFT}\left(\mathbf{y} \odot \mathbf{x}_{\text{ref}}^{*}\right)$, where $\odot$ denotes the Hadamard product and $\mathbf{y} = \begin{bmatrix} y[0] & \hdots & y[N-1] \end{bmatrix}$ and $\mathbf{x}_{\text{ref}} = \begin{bmatrix} x_{\text{ref}}[0] & \hdots & x_{\text{ref}}[N-1] \end{bmatrix}$. Demodulation can be performed by selecting the frequency bin index with the maximum magnitude
\begin{align}
  \hat{s} = \arg\max_{k\in \mathcal{S} } \left( |Y_k| \right). \label{eq:retrieved_symbol_dft}
\end{align}
Using the fast Fourier transform (FFT), the complexity of computing~\eqref{eq:retrieved_symbol_dft} is $O(N\log N)$. These demodulation steps are illustrated in Fig.~\ref{fig:demo_chain}.

\section{Symbol Error Rate Under AWGN} \label{sec:SER_AWGN}

In this section, we first derive the expression for the LoRa SER under additive white Gaussian noise (AWGN), which is useful for later explaining how the SER and the FER can be calculated in the presence of both AWGN and interference.

\subsection{Distribution of the Decision Metric}
In the absence of noise, and with perfect synchronization, the DFT of the dechirped signal $\mathbf{Y}$ has a single frequency bin that contains all the signal energy (i.e., a bin with magnitude $N$) and all remaining $N-1$ bins have zero energy. On the other hand, when AWGN is present, all frequency bins will contain some energy. The distribution of the frequency bin values $Y_k$ for $k \in \mathcal{S}$ is
\begin{align}
  Y_k \sim
  \begin{cases}
    \mathcal{CN}\left(0,2\sigma ^2\right), & k \in \mathcal{S} / s, \\
    \mathcal{CN}\left(N(\cos \phi + j \sin \phi),2\sigma ^2\right), & k = s, \\
  \end{cases}
\end{align}
where $s$ is the transmitted symbol.

Let us define $Y'_k = \frac{Y_k}{\sigma}$ for $k \in \mathcal{S}$. The values $Y'_k$ can be used in~\eqref{eq:retrieved_symbol_dft} instead of $Y_k$ without changing the result and their distribution is
\begin{align}\label{eq:distribution_Yk'}
  Y'_k \sim
  \begin{cases}
    \mathcal{CN}\left(0,2 \right), & k \in \mathcal{S} / s, \\
    \mathcal{CN}\left(\frac{N\cos \phi}{\sigma} + j\frac{N\sin \phi}{\sigma},2 \right), & k = s. \\
  \end{cases}
\end{align}
Thus, using basic properties of the complex normal distribution, we can show that the demodulation metric $|Y'_k|$ follows a Rayleigh distribution for $k \in \mathcal{S} / s$ and a Rice distribution for $k=s$, i.e.,
\begin{align}
  |Y'_k| \sim
  \begin{cases}
    f_{\text{Ra}}(y;1), & k \in \mathcal{S} / s, \\
    f_{\text{Ri}}\left(y;\frac{N}{\sigma},1\right), & k = s.
  \end{cases}
\end{align}

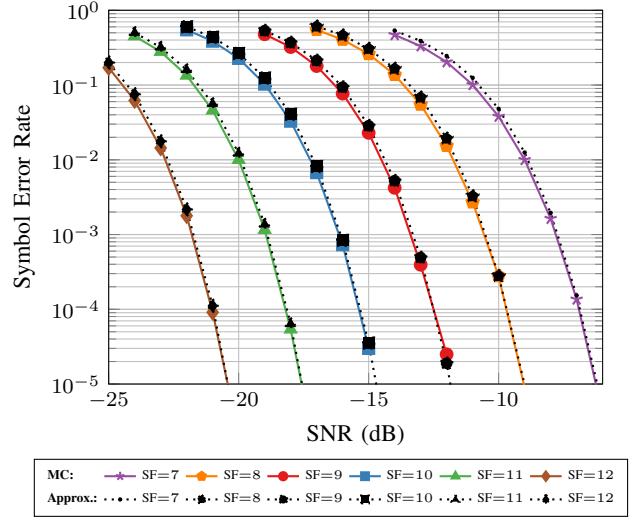
\begin{figure}
  \centering
  \begin{tikzpicture}

	\small

	\begin{semilogyaxis}[
		width = \figurewidth\columnwidth,
		height = \figureheight\columnwidth,
		xlabel = {SNR (dB)},
		ylabel = {Symbol Error Rate},
		label style={font=\small},
    tick label style={font=\footnotesize},
		ylabel near ticks,
		xlabel near ticks,
		xmin = -25, xmax = -6,
		ymin = 1e-5, ymax = 1,
		grid = both,
		legend image post style={scale=0.7},
		legend style={at={(0.45,-0.2)},anchor=north,font=\tiny},
		legend cell align={left},
		legend columns={7},
	]

		\addlegendimage{empty legend}
		\addlegendentry{\textbf{MC:}}

		\addplot[Set1-7-4, thick, solid, mark=star, mark options={scale=1.2}] table[x index=0, y index = 1] {figs/data/RES_SF7_Pi-Inf_OS1_phaseOffset0_symmetries0_0.dat};
		\addlegendentry{SF${=}7$}
		\addplot[Set1-7-5, thick, solid, mark=pentagon*, mark options={scale=1.2}] table[x index=0, y index = 1] {figs/data/RES_SF8_Pi-Inf_OS1_phaseOffset0_symmetries0_0.dat};
		\addlegendentry{SF${=}8$}
		\addplot[Set1-7-1, thick, solid, mark=*, mark options={scale=1.1}] table[x index=0, y index = 1] {figs/data/RES_SF9_Pi-Inf_OS1_phaseOffset0_symmetries0_0.dat};
		\addlegendentry{SF${=}9$}
		\addplot[Set1-7-2, thick, solid, mark=square*, mark options={scale=1.05}] table[x index=0, y index = 1] {figs/data/RES_SF10_Pi-Inf_OS1_phaseOffset0_symmetries0_0.dat};
		\addlegendentry{SF${=}10$}
		\addplot[Set1-7-3, thick, solid, mark=triangle*, mark options={scale=1.2}] table[x index=0, y index = 1] {figs/data/RES_SF11_Pi-Inf_OS1_phaseOffset0_symmetries0_0.dat};
		\addlegendentry{SF${=}11$}
		\addplot[Set1-7-6, thick, solid, mark=diamond*, mark options={scale=1.3}] table[x index=0, y index = 1] {figs/data/RES_SF12_Pi-Inf_OS1_phaseOffset0_symmetries0_0.dat};
		\addlegendentry{SF${=}12$};

		\addlegendimage{empty legend}
		\addlegendentry{\textbf{Approx.:}}
		\addplot[black, thick, dotted, mark=star, mark options={scale=1.2}] table[x index=0, y index = 1] {figs/data/APP_SF7_Pi-Inf_OS1.dat};
		\addlegendentry{SF${=}7$}
		\addplot[black, thick, dotted, mark=pentagon*, mark options={scale=1.2}] table[x index=0, y index = 1] {figs/data/APP_SF8_Pi-Inf_OS1.dat};
		\addlegendentry{SF${=}8$}
		\addplot[black, thick, dotted, mark=*, mark options={scale=1.1}] table[x index=0, y index = 1] {figs/data/APP_SF9_Pi-Inf_OS1.dat};
		\addlegendentry{SF${=}9$}
		\addplot[black, thick, dotted, mark=square*, mark options={scale=1.05}] table[x index=0, y index = 1] {figs/data/APP_SF10_Pi-Inf_OS1.dat};
		\addlegendentry{SF${=}10$}
		\addplot[black, thick, dotted, mark=triangle*, mark options={scale=1.2}] table[x index=0, y index = 1] {figs/data/APP_SF11_Pi-Inf_OS1.dat};
		\addlegendentry{SF${=}11$}
		\addplot[black, thick, dotted, mark=diamond*, mark options={scale=1.3}] table[x index=0, y index = 1] {figs/data/APP_SF12_Pi-Inf_OS1.dat};
		\addlegendentry{SF${=}12$};

	\end{semilogyaxis}

\end{tikzpicture}%
  \caption{Symbol error rate of the LoRa modulation under AWGN for all supported spreading factors $\text{SF} \in \left\{7,\hdots,12\right\}$. Results for Monte Carlo simulations and the approximation in~\eqref{eq:SER_approximation} are shown.}
  \label{fig:serawgn}
\end{figure}

\subsection{Symbol Error Rate}
A symbol error occurs if and only if any of the $|Y'_k|$ values for $k \in \mathcal{S}/s$ exceeds the value of $|Y'_s|$, or, equivalently, if and only if $|Y'_{\max}| > |Y'_s|$, where $|Y'_{\max}| = \max _{k \in \mathcal{S}/s}|Y'_k|$. Using order statistics~\cite{David2003} and the fact that all $|Y'_k|$ for $k \in \mathcal{S}/s$  are i.i.d., the PDF of $|Y'_{\max}|$ can be obtained as
\begin{align} \label{eq:2_to_sf_th_order_pdf}
  f_{|Y'_{\max}|} (y) = \left(N-1\right) f_{\text{Ra}}(y;1) F_{\text{Ra}}\left(y;1\right)^{(N-2)}
\end{align}
Using $f_{|Y'_{\max}|} (y)$, the conditional SER when symbol $s$ is transmitted can be calculated as
\begin{align} \label{eq:formula_SER_only_noise}
  \small
  P(\hat{s} \neq s|s)  & = \int_{y=0}^{+ \infty} \int_{x=0}^{y} f_{\text{Ri}}\left(x;v,1\right) f_{|Y'_{\max}|} (y) {dx} {dy} \\
              & = \int_{y=0}^{+\infty} F_{\text{Ri}}\left(y;v,1\right) f_{|Y'_{\max}|} (y) dy, \label{eq:formula_SER_only_noise_finalform}
\end{align}
with $v = \frac{N}{\sigma}$. The SER for all symbols $s$ is identical, meaning that~\eqref{eq:formula_SER_only_noise_finalform} is in fact equal to the average SER and, if we assume that all symbols are equiprobable, it is also equal to the expected SER.

\subsection{Symbol Error Rate Approximations}
While the evaluation of~\eqref{eq:formula_SER_only_noise_finalform} is in principle straightforward, in practice the values of $N$ in the LoRa modulation are very large so that numerical problems arise. For this reason, two approximations that can be used to efficiently evaluate~\eqref{eq:formula_SER_only_noise_finalform} were derived in~\cite{Elshabrawy2018}. Specifically,~\cite{Elshabrawy2018} used a Gaussian approximation so that $|Y'_s| \mathrel{\dot\sim} \mathcal{N}\left(\frac{N}{\sigma},1\right)$ and $|Y'_{\max}| \mathrel{\dot\sim} \mathcal{N}\left(\mu_{\beta},\sigma^2_{\beta}\right)$ and where appropriate expressions are given to calculate $\mu_{\beta}$ and $\sigma^2_{\beta}$. With our definition of the SNR in \eqref{eq:SNRdef}, the SER can be approximated as
\begin{align}
  P(\hat{s} \neq s) & \approx Q\left(\frac{\sqrt{\text{SNR}} - \left((H_{N-1})^2 - \frac{\pi^2}{12}\right)^{1/4}}{\sqrt{H_{N-1} - \sqrt{(H_{N-1})^2 - \frac{\pi^2}{12}} + 0.5}}\right), \label{eq:SER_approximation}
\end{align}
where $H_n = \sum_{k=1}^n\frac{1}{k}$ denotes the $n$th harmonic number and $Q(\cdot)$ denotes the Q-function. Using several additional approximations~\cite{Elshabrawy2018}, the following more concise version of~\eqref{eq:SER_approximation} is obtained
\begin{align}
  P(\hat{s} \neq s) & \approx Q\left(\sqrt{2\text{SNR}} - \sqrt{2\left(\log(2)\text{SF} + \gamma_{\text{EM}}\right)}\right), \label{eq:SER_approximation_simple}
\end{align}
where $\gamma_{\text{EM}} \approx 0.57722$ is the Euler--Mascheroni constant. We note that it is also possible to directly arrive at~\eqref{eq:SER_approximation_simple} using the methodology of~\cite{Elshabrawy2018} by skipping the intermediate result~\eqref{eq:SER_approximation} and the required additional approximations. It is sufficient to observe that the distribution of the random variable $\hat{\gamma}$ defined in~\cite[Section II-B]{Elshabrawy2018} converges to a Gumbel distribution with $\mu _{\hat{\gamma}} = 2\sigma^2\left(\log(2)\text{SF}+\gamma_{\text{EM}}\right)$ and $\sigma ^2_{\hat{\gamma}} = 4\sigma^2\frac{\pi^2}{6}$ for large $N$ due to the Fisher--Tippett--Gnedenko extremal value theorem~\cite{David2003}.

The SER of the LoRa modulation under AWGN for all supported spreading factors $\text{SF} \in \left\{7,\hdots,12\right\}$ is provided in Fig.~\ref{fig:serawgn}. We show results obtained from both Monte Carlo simulations and the approximation given in~\eqref{eq:SER_approximation}.

\begin{figure}
	\centering
	\includegraphics[width=0.5\textwidth]{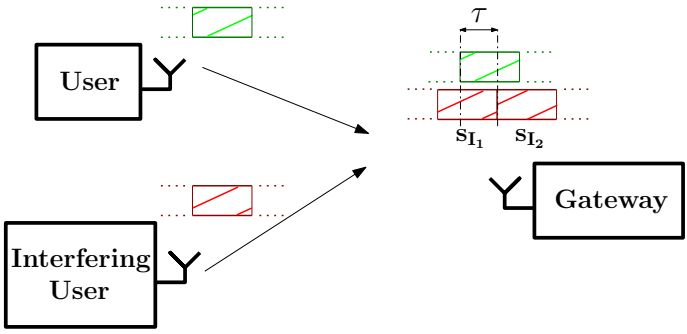}
	\caption{Illustration of LoRa uplink transmission with one interfering user having an arbitrary $\tau$.}
	\label{fig:interf_illustr}
\end{figure}

\section{Symbol Error Rate Under AWGN and \\Same-SF LoRa Interference} \label{sec:SER_interf}

In this section, we analyze the case of a gateway trying to decode the message of a user in the presence of an interfering LoRa device, as depicted in Fig.~\ref{fig:interf_illustr}. This scenario becomes particularly relevant in future deployments with a high density of nodes due to the uncoordinated ALOHA-based random channel access of LoRaWAN~\cite{Georgiou2017}. We assume that the LoRa gateway is perfectly synchronized to the user whose message is decoded. Various synchronization techniques for LoRa have been explained in the literature~\cite{Robyns2018}. It has been shown in~\cite{Croce2018} that interferers with different spreading factors have an average rejection SIR threshold of $-16$~dB while the SIR threshold for same-SF interference is 0~dB. As such, even though the inter-SF interference has a non-negligible effect on the error rate, it is the same-SF interference that has a dominant impact and needs to be modeled first. Therefore, in this work we limit our model to interference signals with the same spreading factor as the one employed by the user of interest. Finally, for simplicity, in this work we only consider one interfering user. In this case, the signal model is
\begin{align}
  y[n] & = hx[n] + h_{I}x_{I}[n] + z[n], \;\; n \in \mathcal{S}, \label{eq:lora_rx_int}
\end{align}
where $h$ is the channel gain between the user of interest and the LoRa gateway, $x[n]$ is the signal of interest, $h_{I}$ is the channel gain between the interferer and the LoRa gateway, $x_{I}[n]$ is the interfering signal, and $z[n] \sim \mathcal{N}(0,\sigma^2)$ is additive white Gaussian noise. Since we assume that $|h| = 1$, the signal-to-interference ratio (SIR) can be defined as
\begin{align}
  \text{SIR}  & = \frac{1}{|h_I|^2} = \frac{1}{P_I},
\end{align}
where we use $P_I$ to denote the power of the interfering user. Since LoRa uses the (non-slotted) ALOHA protocol for medium access control, the interfering signal ${y_{I}[n] = h_{I}x_{I}[n]}$ is not synchronized in any way to the user of interest or the gateway. Due to the lack of synchronization, each LoRa symbol of the user of interest is generally affected by a combination of parts of two distinct interfering LoRa symbols, which we denote by $s_{I_{1}}$ and $s_{I_{2}}$, as shown in Fig.~\ref{fig:interf_illustr}.

Let $\tau$ denote the relative time-offset between the first chip of the symbol of interest $s$ and the first chip of the interfering symbol $s_{I_{2}}$ (i.e., the first chip of the interfering symbol $s_{I_{2}}$ starts $\tau$ chip durations \emph{after} the first chip of $s$). Due to the complete lack of synchronization, we assume that $\tau$ is uniformly distributed in $[0, N)$. We note that in~\cite{Elshabrawy2018b}, the offset $\tau$ is constrained to integer chip durations, which is not particularly realistic since it effectively assumes that the interferer is chip-aligned with the user. Let $ \mathcal{N}_{L_1} = \{0, \dots,\lceil \tau\rceil-1\}$ and $ \mathcal{N}_{L_2} = \{\lceil \tau\rceil, \dots, N-1 \} $. The discrete-time baseband equivalent equation of $x_I[n]$ can be found using~\eqref{eq:LoRa_symbol} for $ s_{I_{1}} $ and $ s_{I_{2}}$, appropriately adjusted to include the offset $\tau$
\begin{align}
x_{I}[n] & =
     \begin{cases}
       e^{j2\pi \left(\frac{(n + N-\tau)^{2}}{2N} + (n + N-\tau)\left(\frac{s_{I_{1}}}{N}-\frac{1}{2}\right)\right)}, & n \in \mathcal{N}_{L_1},\\
       e^{j2\pi \left(\frac{(n - \tau)^{2}}{2N} + (n - \tau)\left(\frac{s_{I_{2}}}{N}-\frac{1}{2}\right)\right)}, & n \in \mathcal{N}_{L_2}.
     \end{cases}
\end{align}
The demodulation of $y[n]$ at the receiver yields
\begin{align}
  \mathbf{Y}  & = \text{DFT}\left(\mathbf{y} \odot \mathbf{x}_{\text{ref}}^{*}\right) \\
              & = \text{DFT}\left(h\mathbf{x} \odot \mathbf{x}_{\text{ref}}^{*}\right) + \text{DFT}\left(h_I\mathbf{x}_{I} \odot \mathbf{x}_{\text{ref}}^{*}\right) + \text{DFT}\left(\mathbf{z} \odot \mathbf{x}_{\text{ref}}^{*}\right).
\end{align}
We call $\text{DFT}\left(\mathbf{x}_{I} \odot \mathbf{x}_{\text{ref}}^{*}\right)$ and $\text{DFT}\left(h_I\mathbf{x}_{I} \odot \mathbf{x}_{\text{ref}}^{*}\right) = \text{DFT}(\mathbf{y}_{I} \odot \mathbf{x}_{\text{ref}}^{*})$ the transmitted and received \emph{interference patterns}, respectively. It is clear that the interference pattern depends on the time-domain interference signal $\mathbf{y}_{I}$, which is in turn a function of the interfering symbols $ s_{I_{1}} $, $ s_{I_{2}} $, the channel $h_I$, and the interferer time-offset $ \tau $.

\begin{figure*}
  \begin{align}
    P\left(\hat{s}\neq s\right) & = 1-\frac{1}{2 \pi N^4}\sum_{s=0}^{N-1}\sum _{s_{I_{1}}=0}^{N-1}\sum _{s_{I_{2}}=0}^{N-1} \int _{\tau = 0}^{N}\int_{\omega=0}^{2 \pi} \int_{y=0}^{+\infty} f_{\text{Ri}}\left(y;v_{s},1\right) \prod_{\substack{k=1\\k\neq s}}^{N} F_{\text{Ri}}(y;v_k,1) dyd\omega d \tau. \label{eq:ser_full}
  \end{align}
  \hrule
\end{figure*}

\subsection{Distribution of the Decision Metric}\label{sec:distdecmetr}
Let $ R_{k} $ denote the value of the transmitted interference pattern at frequency bin $k$, i.e.,
\begin{align}
R_{k} =  \text{DFT}\left(\mathbf{x}_{I} \odot \mathbf{x}_{\text{ref}}^{*}\right)[k], \; k \in \mathcal{S}.
\end{align}

\begin{figure}[t]
	\centering
	\includegraphics[width=0.5\columnwidth]{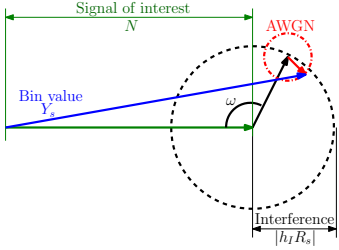}
	\caption{Vector representation of the signal of interest (green), the interference $h_IR_s$ (black), the noise (red), and the bin value $Y_s$ at the bin of interest $s$.}
	\label{fig:binOfInt}
\end{figure}

For a specific combination of a symbol s and an interference pattern $\mathbf{y}_{I}$, adding the interference to the signal of interest corresponds to changing the mean value of the distribution of $ Y'_k $ in~\eqref{eq:distribution_Yk'}, as follows
{
\small
\begin{align}
  Y'_k {\sim}
  \begin{cases}
    \mathcal{CN}\left(\frac{|h_IR_{k}|\cos\theta}{\sigma} {+} j\frac{|h_IR_{k}|\sin\theta}{\sigma},1 \right), & k \in \mathcal{S} / s, \\
    \mathcal{CN}\left(\frac{N\cos\phi {+} |h_IR_{k}|\cos\theta}{\sigma} {+} j\frac{N\sin\phi {+} |h_IR_{k}|\sin\theta}{\sigma},1 \right), & k = s. \\
  \end{cases}
\end{align}
}%
where $\theta = \phase{h_{I}}$ is the phase shift introduced by the interference channel which is fixed for each packet transmission but generally uniformly distributed in $[0,2\pi)$. Thus, in the presence of interference, the demodulation metric $|Y'_k|$ used in~\eqref{eq:retrieved_symbol_dft} is distributed according to
\begin{align}
  |Y'_k| {\sim}
  \begin{cases}
    f_{\text{Ri}}\left(y;\frac{|h_{I}R_k|}{\sigma},1\right), & k \in \mathcal{S} / s, \\
		f_{\text{Ri}}\left(y;\frac{\sqrt{N^2 {+} |h_{I}R_{k}|^2 {+} 2 N|h_{I}R_{k}|\cos(\omega)}}{\sigma},1\right), & k = s, \label{eq:formula_Riceks_noise_interf} \\
  \end{cases}
\end{align}
where we define the phase shift between the user and the interfering user as $\omega = \phi-\theta$ for simplicity. The joint effect of the AWGN and the interference on the signal of interest is illustrated in Fig.~\ref{fig:binOfInt}.
\subsection{Symbol Error Rate}
Similarly to~\eqref{eq:formula_SER_only_noise_finalform}, in the presence of interference, the SER for a given symbol $s$, conditioned on $\mathbf{y}_{I}$ and $\omega$, can be written as
\begin{align} \label{eq:formula_SER_noise_interf}
  \small
	P\left(\hat{s}\neq s|s,\mathbf{y}_{I},\omega\right) & = 1-\int_{y=0}^{+\infty} f_{\text{Ri}}\left(y;v_{s},1\right) F_{|Y'_{\max}|} (y) dy,
\end{align}
where $v_{s} = \frac{1}{\sigma}\sqrt{N^2+ |h_{I}R_{s}|^2 + 2 N|h_{I}R_{s}|\cos(\omega)}$ is the location parameter for the bin $k=s$. The CDF of the $N$th order statistic (i.e., the CDF of the maximum) is known to be $F_{n}(x) = P(X_1 < x) P(X_2 < x) \dots P(X_n < x)$. Due to the conditioning on $\mathbf{y}_{I}$ and $\omega$, each $ |Y'_m| $, for $m\in\left\{1,\hdots,{N}\right\}/ s $, is independent from any other $ |Y'_n| $, for $n\in\left\{1,\hdots,{N}\right\}/ \left\{s,m\right\} $. Thus, we can directly deduce that the CDF of the maximum interfering bin is
\begin{equation}
	F_{|Y'_{\max}|} (y) = \prod_{\substack{k=1\\k\neq s}}^{N} F_{\text{Ri}}(y;v_k,1), \label{eq:cdf_back_bins}
\end{equation}
where $v_k = \nicefrac{|h_{I}R_k|}{\sigma}$. By taking the expectation of $P\left(\hat{s}\neq s|s,\mathbf{y}_{I},\omega\right)$ with respect to $\omega$, we get the SER conditioned on $s$, $\mathbf{y}_{I}$
\begin{align}
	P\left(\hat{s}\neq s|s,\mathbf{y}_{I}\right) & = \frac{1}{2 \pi} \int_{\omega=0}^{2 \pi} P\left(\hat{s}\neq s|s,\mathbf{y}_{I},\omega\right)d\omega \label{eq:cond_prob}.
\end{align}
Recall that, by assumption, $s_{I_{1}}$ and $s_{I_{2}}$ are uniformly distributed in $\mathcal{S}$ and $\tau$ is uniformly distributed in $[0,N)$. As such, the conditional SER (across multiple packets with different time offsets) $P\left(\hat{s}\neq s|s\right)$ can be computed as
\begin{align}
  P\left(\hat{s}\neq s|s\right) & = \frac{1}{N^3}\sum _{s_{I_{1}}=0}^{N-1}\sum _{s_{I_{2}}=0}^{N-1} \int _{0}^{N} P\left(\hat{s}\neq s|s,\mathbf{y}_{I}\right)d \tau. \label{eq:cond_avg}
\end{align}
Finally, $s$ is also uniformly distributed in  $\mathcal{S}$, so that the unconditional SER becomes
\begin{align}
  P\left(\hat{s}\neq s\right) & = \frac{1}{N}\sum_{s=0}^{N-1}P\left(\hat{s}\neq s|s\right). \label{eq:final_avg}
\end{align}
The full expression for $P\left(\hat{s}\neq s\right) $ is given in~\eqref{eq:ser_full}.

\section{Computational Complexity Reduction} \label{sec:symmetries}

Apart from the numerical problems that arise from the product of $(N{-}1)$ CDFs in~\eqref{eq:ser_full}, an additional practical issue is that the computational complexity of evaluating~\eqref{eq:ser_full} is very high. In particular, the complexity of computing the three sums scales like $N^3$ and three integrals need to be numerically evaluated in order to obtain each of the $N^3$ summation terms. In this section we show that there exist sets of~\emph{equivalent interference patterns} that can be exploited in order to reduce the complexity of evaluating~\eqref{eq:ser_full}.

\subsection{Interference Patterns}
We first derive an explicit form for the magnitude of the transmitted interference pattern $R_k,~k\in\left\{0,\hdots,{N-1}\right\}$, which we will then use to show the existence of equivalent interference patterns. Note that the offset $\tau$ can be split into an integer part $L$ and a non-integer part $\lambda$
\begin{align}
  L & = \left \lfloor{\tau}\right \rfloor,\\
  \lambda & = \tau - \left \lfloor{\tau}\right \rfloor,
\end{align}
where $ L $ and $\lambda$ correspond to the inter-chip and intra-chip misalignments between the user and the interferer, respectively.

Using the definition of the DFT and after some algebraic transformations, we have
\begin{equation}
  R_{k} = \sum_{n=0}^{N-1} Z_{k,n},
\end{equation}
where $Z_{k,n}$ is defined as
\begin{align}\label{eq:Zeta}
Z_{k,n} &=
     \begin{cases}
      T_{1} e^{j\frac{2\pi}{N} n(s_{I_{1}}-k-\tau)} e^{-j2\pi \lambda}, & n \in \mathcal{N}_{L_1},\\
      T_{2} e^{j\frac{2\pi}{N} n(s_{I_{2}}-k-\tau)}, & n \in \mathcal{N}_{L_2},
     \end{cases}
\end{align}
and $T_{i}$ are terms that are independent of the summation variable $n$ which are given by
\begin{align}
  T_{i} &= e^{j2\pi \frac{\tau^2}{2N}} e^{j2\pi \frac{\tau}{2}} e^{-j2\pi \frac{s_{I_{i}} \tau}{N}}.
\end{align}
Using the geometric series sum formula and after some relatively straightforward operations, $ R_{k} $ can be written as
\begin{align}
  R_{k} & = A_{k,1} e^{-j\theta_{k,1}} + A_{k,2} e^{-j\theta_{k,2}}, \label{eq:short_A_theta}
\end{align}%
where
\begin{align}
	A_{k,1} & = \frac{\sin \left( \frac{\pi}{N} (s_{I_{1}}-k-\tau)\lceil \tau \rceil \right)}{\sin \left( \frac{\pi}{N} (s_{I_{1}}-k-\tau) \right)}, \label{eq:A1} \\
  A_{k,2} &=  \frac{\sin \left( \frac{\pi}{N} (s_{I_{2}}-k-\tau)(N-\lceil \tau \rceil) \right)}{\sin \left( \frac{\pi}{N} (s_{I_{2}}-k-\tau) \right)}, \label{eq:A2}
\end{align}
and
\begin{align}
  \theta_{k,1} & = \frac{\pi}{N} \left( {-}\tau^2 + (\lambda{-}L)N + s_{I_{1}}(2\tau -\lceil \tau \rceil +1) + \right. \nonumber\\
             &  + \left. k(\lceil \tau \rceil -1) {+} \tau(\lceil \tau \rceil -1) \right), \\
  \theta_{k,2} & = \frac{\pi}{N} \left( {-}\tau^2 + s_{I_{2}}(2\tau -\lceil \tau \rceil +1-N) + \right. \nonumber\\
   &  + \left. k(\lceil \tau \rceil -1+N) + \tau(\lceil \tau \rceil -1) \right).
\end{align}
For the special case where $\tau$ is an integer and $k = [s_{I_{1}}-\tau]_N$ and $k = [s_{I_{2}}-\tau]_N$, \eqref{eq:A1} and \eqref{eq:A2}, respectively, are of the indeterminate form $\frac{0}{0}$. Using L'H\^{o}pital's rule, it can be shown that in these cases we have $ A_{k,1} = \lceil \tau \rceil $, and $ A_{k,2} = N - \lceil \tau \rceil $. Using Euler's formula, and for all $k \in \mathcal{S},$ the magnitude of $R_k$ in~\eqref{eq:short_A_theta} can be written as
\begin{align}
  |R_{k}| & = \sqrt{A_{k,1}^{2} + A_{k,2}^{2} + 2A_{k,1}A_{k,2}\cos(\theta_{k,1} - \theta_{k,2})}. \label{eq:final_Yk}
\end{align}

\subsection{Equivalent Interference Patterns}
We first give a definition for the equivalent interference patterns. Then, we show two equivalent interference pattern properties and we explain how they can be used in order to reduce the computational complexity of evaluating~\eqref{eq:ser_full}.
\begin{definition}
An interference pattern $\mathbf{y}_{I_1}$ is said to be equivalent with respect to some other interference pattern $\mathbf{y}_{I_2}$ if it contains exactly the same set of frequency bin magnitudes $|R_{k}|$, $ k \in \mathcal{S} $, irrespective of the order of these magnitudes within the set.%
\end{definition}
We note that the ordering of the magnitudes $|R_{k}|$ does not change the distribution of $ |Y'_{\max}|$, thus the probability of $|Y'_{\max}| > |Y'_s|$ is not affected. Therefore, equivalent interference patterns result in exactly the same conditional SER $P(\hat{s}\neq s|s,\mathbf{y}_{I})$, meaning that it is sufficient to compute each distinct interference pattern once for the evaluation of the unconditional SER $P(\hat{s}\neq s)$ given in~\eqref{eq:ser_full}. Naturally, care has to be taken so that the contribution of each distinct interference pattern is weighted according to how many other equivalent interference patterns exist.

\begin{prop}\label{prop:Prop_delta}
Let $\delta \in \{0,1,...,N-1\}$ and $s_{I_1} \geq s_{I_2}$ without loss of generality and let $\tau$ be fixed. Moreover, let $s'_{I_1} = [s_{I_1}+\delta]_N$ and $s'_{I_2} = [s_{I_2}+\delta]_N$. Then there exist the following two sets of equivalent interference patterns
\begin{align}
  \mathcal{Y}_{I_1} & = \left\{ \mathbf{y}_{I}(s'_{I_1} ,s'_{I_2} ,\tau):  s'_{I_1} \geq s'_{I_2} \right\}, \label{eq:y1}\\
  \mathcal{Y}_{I_2} & = \left\{ \mathbf{y}_{I}(s'_{I_1} ,s'_{I_2} ,\tau):  s'_{I_1} < s'_{I_2} \right\}, \label{eq:y2}
\end{align}
where the interference patterns in $\mathcal{Y}_{I_1}$ are generally not equivalent versions of the patterns in $\mathcal{Y}_{I_2}$. Furthermore, the cardinalities of the two sets are
\begin{align}
  \left|\mathcal{Y}_{I_1}\right| & = N-(s_{I_1}-s_{I_2}), \label{eq:card1}\\
  \left|\mathcal{Y}_{I_2}\right| & = (s_{I_1}-s_{I_2}). \label{eq:card2}
\end{align}
In the special case where $\lambda = 0$ (i.e., when $\tau$ is an integer), all interference patterns in both $\mathcal{Y}_{I_1}$ and $\mathcal{Y}_{I_2}$ are equivalent.
\end{prop}%
\begin{IEEEproof}
A detailed proof is given in the Appendix.
\end{IEEEproof}

\begin{prop}\label{prop:Prop_offsets}
Let $\tau \in \left[0,N-1\right)$ and let $\tau'$ be
\begin{align}
  \tau ' & = (N-1)-\tau. \label{eq:deltalprime}
\end{align}
Then, the interference patterns $\mathbf{y}_{I}(s_{I_1}, s_{I_2} ,\tau)$ and $\mathbf{y}_{I}(s_{I_1}, s_{I_2} ,\tau')$ are equivalent.
\end{prop}%
\begin{IEEEproof}
A detailed proof is given in the Appendix.
\end{IEEEproof}

\subsection{Complexity Reduction}
The essence of Proposition~\ref{prop:Prop_delta} is that there are only two sets of distinct interference patterns for each value of $s_I = [s_{I_1}{-}s_{I_2}]_{N}$. Let $P_e(\mathcal{Y}_{I_{i}}) = P\left(\hat{s}\neq s|s,\mathbf{y}_{I_i}(s_{I},\tau)\right)$, where $\mathbf{y}_{I_i}(s_I,\tau)$ denotes any (equivalent) element of $\mathcal{Y}_{I_{i}}$. Then, the double sum in~\eqref{eq:cond_avg} can be simplified to
\begin{align}
  P\left(\hat{s}\neq s|s\right) & = \frac{1}{N^3}\sum _{s_{I_{1}}=0}^{N-1}\sum _{s_{I_{2}}=0}^{N-1} \int _{0}^{N} P\left(\hat{s}\neq s|s,\mathbf{y}_{I}\right)d \tau \\
                                  & = \frac{1}{N^2}\sum _{s_{I}=0}^{N{-}1} \int _{0}^{N} \left(\frac{1}{N}\sum_{i=1}^2\left|\mathcal{Y}_{I_i}\right|P_e(\mathcal{Y}_{I_{i}})\right)d \tau, \label{eq:simpl1}
\end{align}
which reduces the complexity of evaluating~\eqref{eq:ser_full} by a factor of $N/2$. In the special case of integer offsets $\tau$ (i.e., $\lambda = 0$ and $\tau = L$) the above integral can be simplified to a sum over all the integer offsets $L$. Moreover, in such a case, the sets $\mathcal{Y}_{I_i}$ are equivalent, meaning that there is only one set of equivalent interference patterns $\mathcal{Y}_{I}$. Therefore, the above expression can be further simplified to the expression found in~\cite{Elshabrawy2018b}
\begin{align}
    P\left(\hat{s}\neq s|s\right) & = \frac{1}{N^2}\sum _{s_{I}=0}^{N{-}1} \sum_{L=0}^{{N{-}1}} P_e(\mathcal{Y}_I).
\end{align}
Proposition~\ref{prop:Prop_offsets} essentially means that any interference pattern with $\tau \in (\nicefrac{(N{-}1)}{2},N{-}1]$ is equivalent with exactly one interference pattern with $\tau \in [0,\nicefrac{(N{-}1)}{2})$. It is important to note that we have not shown any property about the interference patterns in the region $\tau \in (N-1,N)$, so we consider this region separately.\footnote{It is relatively simple to verify that the region $\tau \in (N-1,N)$ contains equivalent interference patterns around the point $\tau = N-\frac{1}{2}$ following the syllogism of the proof of Property~\ref{prop:Prop_offsets}. However, this property only marginally reduces the computational complexity and we thus neither use it nor prove it explicitly.} If we let $\tilde{P}_{e} = \left(\frac{1}{N}\sum_{i=1}^2\left|\mathcal{Y}_{I_i}\right|P_e(\mathcal{Y}_{I_{i}})\right)$, the expression in \eqref{eq:simpl1} can be re-written as
\begin{align}
  P\left(\hat{s}\neq s|s\right) & = \frac{1}{N^2}\sum _{s_{I}=0}^{N{-}1} \left(2\int _{0}^{\frac{N{-}1}{2}} \tilde{P}_{e}d \tau + \int _{N{-}1}^N \tilde{P}_{e}d \tau\right), \label{eq:int1}
\end{align}
which reduces the complexity of evaluating~\eqref{eq:ser_full} by an additional factor of approximately $2$. In the special case of integer offsets $\tau$ (i.e., $\lambda = 0$ and $\tau = L$) any interference pattern with $\tau \in \left\{\nicefrac{N}{2},N{-}1\right\}$ is equivalent with exactly one interference pattern with $\tau \in \left\{0,\nicefrac{N}{2}{-}1\right\}$. Therefore, the above two integrals can be simplified to a summation over all the integer offsets $L$. In this integer-offset case we have
\begin{align}
    P\left(\hat{s}\neq s|s\right) & = \frac{1}{N\left(\frac{N}{2}\right)}\sum _{s_{I}=0}^{N{-}1} \sum_{L=0}^{\frac{N}{2}{-}1} P_e(\mathcal{Y}_I). \label{eq:simpl2}
\end{align}
We note that the corresponding simplification that is used in~\cite{Elshabrawy2018b}, corresponding to the special chip-aligned case, is different than the simplification we gave in~\eqref{eq:simpl2}. Specifically, in~\cite{Elshabrawy2018b} the upper limit of the sum over $L$ is $N/2$ and, consequently, the normalization is done with $\left(\frac{N}{2}+1\right)$ instead of $\frac{N}{2}$. However, the interference pattern resulting from $\tau = \frac{N}{2}$ is equivalent with the interference pattern resulting from $\tau = \frac{N}{2}{-}1$, which has already been considered in the sum. 

\subsection{Symbol Error Rate Approximation} \label{sec:SER_approx}
Since, even with the above simplifications, the complexity of evaluating~\eqref{eq:ser_full} is very high, we derive a low-complexity approximation for~\eqref{eq:ser_full}. Using the triangle inequality, we can simplify~\eqref{eq:final_Yk} to
\begin{align}
  |R_{k}| & \approx  |A_{k,1}|+|A_{k,2}|. \label{eq:triangle}
\end{align}
With this simplification, the $A_{k,1}A_{k,2}\cos(\theta_{k,1}-\theta_{k,2})$ term that leads to the existence of two sets of interference patterns $\mathcal{Y}_{I_{1}}$ and $\mathcal{Y}_{I_{2}}$ (cf. proof of Proposition~\ref{prop:Prop_delta}) disappears.
Thus, there is only a single set of interference patterns $\mathcal{Y}_{I}$ and~\eqref{eq:simpl1} can be simplified to
\begin{align}
  P\left(\hat{s}\neq s|s\right) & \approx \frac{1}{N^2}\sum _{s_{I}=0}^{N{-}1} \int _{0}^{N} P_e(\mathcal{Y}_{I})d \tau.
\end{align}
Moreover, we also approximate~\eqref{eq:int1} by ignoring the second integral for $\tau \in (N-1,N)$ so that
\begin{align}
  P\left(\hat{s}\neq s|s\right) & \approx \frac{2}{N^2}\sum _{s_{I}=0}^{N{-}1} \int _{0}^{\frac{N{-}1}{2}} P_e(\mathcal{Y}_{I})d \tau, \label{eq:intapprox}
\end{align}
We now follow a procedure that is similar to the procedure in~\cite{Elshabrawy2018b} in order to derive a simple approximation for $P_e(\mathcal{Y}_{I})$. First, we assume that the interference-induced SER is dominated  by the maximum of $|R_k|$. Thus, we are interested in evaluating
\begin{align}
  |R_{k_{\max}}| & = \max_k\left(|A_{k,1}|+|A_{k,2}|\right).
\end{align}
Without loss of generality, we assume that $s_{I_2}=0$, so that $s_I = s_{I_1}$. Since $\tau \in \left[0,\nicefrac{(N-1)}{2}\right)$ and due to~\eqref{eq:A1} and~\eqref{eq:A2} it holds that $\max_k\left(|A_{k,2}|\right)> \max_k\left(|A_{k,1}|\right)$. Based on this observation, we choose
\begin{align}
  k_{\max} & \approx \arg \max_k \left(|A_{k,2}|\right) = \lfloor \tau \rceil,
\end{align}
so that we can easily approximate $|R_{k_{\max}}|$ as
\begin{align}
    |R_{k_{\max}}| & \approx |A_{\lfloor \tau \rceil,1} + A_{\lfloor \tau \rceil,2}|. \label{eq:maxapprox}
\end{align}
The probability of the event that the (maximum) interference bin $\lfloor \tau \rceil$ coincides with the bin of the signal-of-interest $s$ is $\frac{1}{N}$. Since in LoRa $N$ is relatively large ($N > 2^7$), the impact of the aforementioned event on the total error probability is negligible, and therefore, for the approximation of the SER, we only consider the cases where $\lfloor \tau \rceil \neq s$. The aforementioned fact, combined with~\eqref{eq:maxapprox} which says that all bins except $s$ and $\lfloor \tau \rceil$ are zero-valued, means that the approximation of the SER does not depend on the value of $s$. As such, $P\left(\hat{s}\neq s|s\right)= P\left(\hat{s}\neq s\right)$ and calculating the expectation over $s$ can be avoided. Only considering $\lfloor \tau \rceil \neq s$ also has the convenient side-effect that we ignore the only case of~\eqref{eq:formula_Riceks_noise_interf} which contains $\omega$, meaning that we can entirely avoid the integration over $\omega$ in the computation of $P_e(\mathcal{Y}_{I})$. Let $P^{(I)}(\hat{s}\neq s)$ denote the interference-dominated SER resulting from the approximation in~\eqref{eq:maxapprox}. As explained in Section~\ref{sec:distdecmetr}, $|Y'_{k_{\max}}|$ follows a Rice distribution, which can be approximated by a Gaussian distribution for large location parameters~\cite{Elshabrawy2018b} so that
\begin{align}
  |Y'_{k_{\max}}| \mathrel{\dot\sim} \mathcal{N}\left(\frac{|h_I||R_{k_{\max}}|}{\sigma},1\right).
\end{align}
Using the Gaussian approximation, the interference-dominated SER $P^{(I)}(\hat{s}{\neq} s)$ can be computed as
\begin{align}
  P^{(I)}(\hat{s}{\neq} s)  & = \frac{1}{N\left(\frac{N}{2}\right)}\sum _{s_{I}=0}^{N{-}1} \int _{0}^{\frac{N{-}1}{2}} Q\left(\frac{N-|h_I||R_{k_{\max}}|}{\sqrt{2\sigma^2}}\right)d \tau, \label{eq:approxint}
\end{align}
where $Q(\cdot)$ denotes the Q-function and the integral can be evaluated numerically by discretizing the interval $[0,\nicefrac{(N{-}1)}{2})$ with a step size $\epsilon$ as
\begin{align}
  P^{(I)}(\hat{s}{\neq} s)  & \approx \frac{\epsilon}{N\left(\frac{N}{2}\right)}\sum _{s_{I}=0}^{N{-}1} \sum _{\tau \in \mathcal{T}} Q\left(\frac{N-|h_I||R_{k_{\max}}|}{\sqrt{2\sigma^2}}\right), \label{eq:approxint2}
\end{align}
where $\mathcal{T} = \left\{0,\epsilon,2\epsilon,\hdots,\frac{N-1}{2}-\epsilon\right\}$.

We note that, in the low SNR (i.e., AWGN-limited) regime, the above approximation becomes inaccurate, since all bins have similar values and no single bin dominates the error rate. Let $P^{(N)}(\hat{s}\neq s)$ denote the SER under AWGN given in~\eqref{eq:formula_SER_only_noise_finalform} (which can be evaluated efficiently using the approximation in~\eqref{eq:SER_approximation}). Then, a final estimate of the SER that is more accurate also in the low SNR regime~\cite{Elshabrawy2018b} can be obtained as
\begin{align}
    P(\hat{s}{\neq} s)  & \approx P^{(N)}(\hat{s}{\neq} s) {+} \left(1{-}P^{(N)}(\hat{s}{\neq} s)\right)P^{(I)}(\hat{s}{\neq} s). \label{eq:approxfinal}
\end{align}

\section{Frame Error Rate Under AWGN and\\Same-SF Interference}\label{sec:fer}
Since network simulators, such as the ones presented in~\cite{Abeele2017,Reynders2018,Bor2016,Pop2017}, typically operate on a frame level, the frame error rate (FER) is generally of greater practical relevance than the SER. The expression for the SER derived in Section~\ref{sec:SER_interf} can not be used directly to evaluate the FER because it includes an expectation over $\tau$ and an expectation over $\omega$, while all symbols in a frame would experience the same $\tau$ and $\omega$. For this reason, in this section we derive an expression to approximate the FER. We note that we are considering the FER of an uncoded system. Thus, the expression we derive can be used for the LoRa mode that uses a channel code of rate $\nicefrac{4}{5}$, which has error-detection but no error-correction capabilities.

We first make the following simplifying assumptions. We assume that perfect frame synchronization for the user of interest is achieved even in the presence of interference. Then, we also assume that the interfering frame has the same length as the frame of interest. The latter assumption is taken only for clarity of the presentation and does not restrict our results from being easily generalized to any interfering frame length. Due to the time offset between the frames, only part of the frame of interest is affected by interference.

Let the vector $\mathbf{s}$ denote a frame of $F$ LoRa symbols and let the vector $\mathbf{\hat{s}}$ denote the estimated frame at the receiver. The FER can then be defined as $P(\hat{\mathbf{s}}\neq \mathbf{s})$. Moreover, let $F_{i}$, where $ i \in \{1, \dots, F \} $, denote the number of symbols in the frame that are affected by the interfering frame. The value of $F_{i}$ depends on the relative position of the two frames. As mentioned in~\cite{Croce2018}, the random relative position of the two frames plays an important role on the final FER. We consider the final FER as the expectation over all the possible relative positions of the two frames. We note that, except in the case of perfect alignment between the frame of interest and the interference, there always exists one symbol that is only partially affected by interference. For simplicity, we approximate this situation by considering the partially-affected symbol as fully-affected by interference, thus including it in $F_{i}$. The number of symbols in the frame that are affected only by AWGN is $F - F_{i}$. Since the frame of interest and the interfering frame have the same length,
the number $F_{i}$ of interfered symbols in a frame is uniformly distributed in $ \{1, \dots, F \} $.

Since all symbols of the interfering frame have the same offset $\tau$, the probability, for a given $ F_{i} $, that all $F_{i}$ symbols under interference and all $ F - F_{i}$ symbols under AWGN are correct is 
{\normalsize
\begin{align}
 P(\hat{s} = s|F_{i},\tau) & = \left(1- P(\hat{s}\neq s|\tau)\right)^{F_{i}}\left(1-P^{(N)}(\hat{s}\neq s)\right)^{F-F_{i}},
\end{align}
}
where $P(\hat{s}\neq s|\tau)$ can be approximated as
\begin{align}
  P(\hat{s}\neq s|\tau) & = \frac{1}{N}\sum _{s_{I}=0}^{N{-}1} Q\left(\frac{N-|h_I||R_{k_{\max}}|}{\sqrt{2\sigma^2}}\right).
\end{align}
By using a similar simplification to~\eqref{eq:intapprox}, the conditional frame error rate $P(\hat{\mathbf{s}}\neq \mathbf{s}|F_{i})$ can be approximated as
\begin{align}
    P(\hat{\mathbf{s}} \neq  \mathbf{s}|F_{i})  & \approx \frac{2}{N}\int _{0}^{\frac{N-1}{2}}\left(1-P(\hat{s} = s|F_{i},\tau)\right) d\tau.
\end{align}
Finally, we take the expectation over all possible values of $ F_{i} $ and we obtain the final expression for the FER
\begin{align}
  P(\hat{\mathbf{s}} \neq  \mathbf{s})  & \approx \frac{1}{F}\sum _{F_{i}=1}^{F} P(\hat{\mathbf{s}} \neq  \mathbf{s}|F_{i}).\label{eq:approxFERfinal}
\end{align} 

\section{Results} \label{sec:results}

In this section, we provide numerical results for the SER and the FER of a LoRa user with same-SF interference.

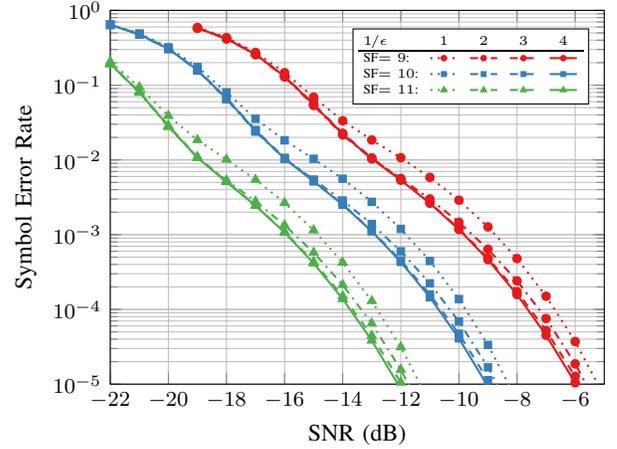
\begin{figure}[t]
  \centering
  \begin{tikzpicture}

	\small

	\begin{semilogyaxis}[
		width = \figurewidth\columnwidth,
		height = \figureheight\columnwidth,
		xlabel = {SNR (dB)},
		ylabel = {Symbol Error Rate},
		label style={font=\small},
    tick label style={font=\footnotesize},
		ylabel near ticks,
		xlabel near ticks,
		xmin = -22, xmax = -5,
		ymin = 1e-5, ymax = 1,
		grid = both,
		legend image post style={scale=0.6},
	]


		\addplot[Set1-7-1, thick, dotted, mark=*, mark options={scale=0.7, solid}] table[x index=0, y index = 1] {figs/data/APP_SF9_Pi-3.00_OS1.dat};
		\label{SF9APPNCNOS1}
		\addplot[Set1-7-1, thick, dashdotted, mark=*, mark options={scale=0.7, solid}] table[x index=0, y index = 1] {figs/data/APP_SF9_Pi-3.00_OS2.dat};
		\label{SF9APPNCNOS2}
		\addplot[Set1-7-1, thick, dashed, mark=*, mark options={scale=0.7, solid}] table[x index=0, y index = 1] {figs/data/APP_SF9_Pi-3.00_OS3.dat};
		\label{SF9APPNCNOS3}	
		\addplot[Set1-7-1, thick, solid, mark=*, mark options={scale=0.7, solid}] table[x index=0, y index = 1] {figs/data/APP_SF9_Pi-3.00_OS4.dat};
		\label{SF9APPNCNOS4}

		\addplot[Set1-7-2, thick, dotted, mark=square*, mark options={scale=0.6, solid}] table[x index=0, y index = 1] {figs/data/APP_SF10_Pi-3.00_OS1.dat};
		\label{SF10APPNCNOS1}
		\addplot[Set1-7-2, thick, dashdotted, mark=square*, mark options={scale=0.6, solid}] table[x index=0, y index = 1] {figs/data/APP_SF10_Pi-3.00_OS2.dat};
		\label{SF10APPNCNOS2}
		\addplot[Set1-7-2, thick, dashed, mark=square*, mark options={scale=0.6, solid}] table[x index=0, y index = 1] {figs/data/APP_SF10_Pi-3.00_OS3.dat};
		\label{SF10APPNCNOS3}	
		\addplot[Set1-7-2, thick, solid, mark=square*, mark options={scale=0.6, solid}] table[x index=0, y index = 1] {figs/data/APP_SF10_Pi-3.00_OS4.dat};
		\label{SF10APPNCNOS4}

		\addplot[Set1-7-3, thick, dotted, mark=triangle*, mark options={scale=0.8, solid}] table[x index=0, y index = 1] {figs/data/APP_SF11_Pi-3.00_OS1.dat};
		\label{SF11APPNCNOS1}
		\addplot[Set1-7-3, thick, dashdotted, mark=triangle*, mark options={scale=0.8, solid}] table[x index=0, y index = 1] {figs/data/APP_SF11_Pi-3.00_OS2.dat};
		\label{SF11APPNCNOS2}
		\addplot[Set1-7-3, thick, dashed, mark=triangle*, mark options={scale=0.8, solid}] table[x index=0, y index = 1] {figs/data/APP_SF11_Pi-3.00_OS3.dat};
		\label{SF11APPNCNOS3}	
		\addplot[Set1-7-3, thick, solid, mark=triangle*, mark options={scale=0.8, solid}] table[x index=0, y index = 1] {figs/data/APP_SF11_Pi-3.00_OS4.dat};
		\label{SF11APPNCNOS4}

		\node [draw,fill=white,inner sep=2pt] at (rel axis cs: 0.73,0.85) {
		\tiny
		\begin{tabular}{lcccc}
			$\nicefrac{1}{\epsilon}$	& $1$ 								& $2$ 								& $3$ 								& $4$ \\
			\hline	
			SF$=9$:	 									& \ref{SF9APPNCNOS1} 	&	\ref{SF9APPNCNOS2} 	&	\ref{SF9APPNCNOS3} 	&	\ref{SF9APPNCNOS4} \\
			SF$=10$: 									& \ref{SF10APPNCNOS1} & \ref{SF10APPNCNOS2}	& \ref{SF10APPNCNOS3}	& \ref{SF10APPNCNOS4} \\
			SF$=11$: 									& \ref{SF11APPNCNOS1} & \ref{SF11APPNCNOS2}	& \ref{SF11APPNCNOS3}	& \ref{SF11APPNCNOS4}
		\end{tabular}};

	\end{semilogyaxis}

\end{tikzpicture}%
  \caption{Symbol error rate approximation of the LoRa modulation under AWGN and same-SF interference for $\text{SF}
\in \left\{9,10,11\right\}$ and $P_I = {-}3$ dB, for various values of the oversampling factor $\nicefrac{1}{\epsilon}$.}
  \label{fig:serintepsilon}
\end{figure}

\subsection{Choice of Discretization Step}
In Fig.~\ref{fig:serintepsilon} we show the evaluation of \eqref{eq:approxint2} with different discretization steps $\epsilon$ (or, equivalently, oversampling factors $\nicefrac{1}{\epsilon}$). We observe that very small gains in accuracy are obtained after $\epsilon = \nicefrac{1}{3}$. As such, in the remainder of this section we use $\epsilon = \nicefrac{1}{5}$ to be on the safe side. This choice also means that the complexity of evaluating~\eqref{eq:approxint2} is not significantly higher than the complexity of evaluating the corresponding approximation in~\cite{Elshabrawy2018b} (which is obtained for $\epsilon = 1$). In general, the complexity of evaluating~\eqref{eq:approxint2} is $\nicefrac{1}{\epsilon}$ times higher than the complexity of evaluating the corresponding approximation in~\cite{Elshabrawy2018b}, but still very low compared to simulation-based studies. We note that for the Monte Carlo simulations we use a more conservative $\epsilon = \nicefrac{1}{10}$, since they are used as a comparison baseline.

\subsection{Symbol Error Rate}
In Fig.~\ref{fig:serintphase}, we show the results of a Monte Carlo simulation for the SER of a LoRa user for all possible spreading factors $\text{SF} \in \left\{7,\hdots,12\right\}$, under the effect of same-SF interference with an SIR of $3$ dB (i.e., $P_I=-3$ dB) and AWGN. The SER when there is only AWGN is also included in the figure with thick transparent lines (taken from Fig.~\ref{fig:serawgn}). We can clearly observe the strong impact of the interference on the SER when comparing to the case where there is only AWGN. The black dotted lines in the figure depict the SER when the relative phase offset $\omega$ between the interferer and the user is not taken into account in the Monte Carlo simulation. It is interesting to observe that $\omega$ does not seem to play an important role for the SER, which further justifies ignoring $\omega$ in the approximation of Section~\ref{sec:SER_approx}.

\begin{figure}[t]
  \centering
  \begin{tikzpicture}

	\small

	\begin{semilogyaxis}[
		width = \figurewidth\columnwidth,
		height = \figureheight\columnwidth,
		xlabel = {SNR (dB)},
		ylabel = {Symbol Error Rate},
		label style={font=\small},
    tick label style={font=\footnotesize},
		ylabel near ticks,
		xlabel near ticks,
		xmin = -25, xmax = 0,
		ymin = 1e-5, ymax = 1,
		grid = both,
		legend image post style={scale=0.7},
		legend style={at={(0.45,-0.2)},anchor=north,font=\tiny},
		legend cell align={left},
		legend columns={7},
	]

		\addlegendimage{empty legend}
    \addlegendentry{\textbf{PO:}} 

		\addplot[Set1-7-5, thick, solid, mark=star, mark options={scale=1.2}] table[x index=0, y index = 1] {figs/data/RES_SF7_Pi-3.00_OS10_phaseOffset1_symmetries0_0.dat};
		\addlegendentry{SF${=}7$}
		\addplot[Set1-7-6, thick, solid, mark=pentagon*, mark options={scale=1.2}] table[x index=0, y index = 1] {figs/data/RES_SF8_Pi-3.00_OS10_phaseOffset1_symmetries0_0.dat};
		\addlegendentry{SF${=}8$}
		\addplot[Set1-7-1, thick, solid, mark=*, mark options={scale=1.1}] table[x index=0, y index = 1] {figs/data/RES_SF9_Pi-3.00_OS10_phaseOffset1_symmetries0_0.dat};
		\addlegendentry{SF${=}9$}
		\addplot[Set1-7-2, thick, solid, mark=square*, mark options={scale=1.05}] table[x index=0, y index = 1] {figs/data/RES_SF10_Pi-3.00_OS10_phaseOffset1_symmetries0_0.dat};
		\addlegendentry{SF${=}10$}
		\addplot[Set1-7-3, thick, solid, mark=triangle*, mark options={scale=1.2}] table[x index=0, y index = 1] {figs/data/RES_SF11_Pi-3.00_OS10_phaseOffset1_symmetries0_0.dat};
		\addlegendentry{SF${=}11$}
		\addplot[Set1-7-4, thick, solid, mark=diamond*, mark options={scale=1.3}] table[x index=0, y index = 1] {figs/data/RES_SF12_Pi-3.00_OS10_phaseOffset1_symmetries0_0.dat};
		\addlegendentry{SF${=}12$};

		\addlegendimage{empty legend}
    \addlegendentry{\textbf{Non-PO:}} 
		\addplot[black, thick, dotted, mark=star, mark options={scale=0.6, solid}] table[x index=0, y index = 1] {figs/data/RES_SF7_Pi-3.00_OS10_phaseOffset0_symmetries0_0.dat};
		\addlegendentry{SF${=}7$}
		\addplot[black, thick, dotted, mark=pentagon*, mark options={scale=0.5, solid}] table[x index=0, y index = 1] {figs/data/RES_SF8_Pi-3.00_OS10_phaseOffset0_symmetries0_0.dat};
		\addlegendentry{SF${=}8$}
		\addplot[black, thick, dotted, mark=*, mark options={scale=0.6, solid}] table[x index=0, y index = 1] {figs/data/RES_SF9_Pi-3.00_OS10_phaseOffset0_symmetries0_0.dat};
		\addlegendentry{SF${=}9$}
		\addplot[black, thick, dotted, mark=square*, mark options={scale=0.6, solid}] table[x index=0, y index = 1] {figs/data/RES_SF10_Pi-3.00_OS10_phaseOffset0_symmetries0_0.dat};
		\addlegendentry{SF${=}10$}
		\addplot[black, thick, dotted, mark=triangle*, mark options={scale=0.6, solid}] table[x index=0, y index = 1] {figs/data/RES_SF11_Pi-3.00_OS10_phaseOffset0_symmetries0_0.dat};
		\addlegendentry{SF${=}11$}
		\addplot[black, thick, dotted, mark=diamond*, mark options={scale=0.6, solid}] table[x index=0, y index = 1] {figs/data/RES_SF12_Pi-3.00_OS10_phaseOffset0_symmetries0_0.dat};
		\addlegendentry{SF${=}12$}

		\addplot[black, ultra thick, solid, opacity=0.25] table[x index=0, y index = 1] {figs/data/RES_SF7_Pi-Inf_OS1_phaseOffset0_symmetries0_0.dat};
		\addplot[black, ultra thick, solid, opacity=0.25] table[x index=0, y index = 1] {figs/data/RES_SF8_Pi-Inf_OS1_phaseOffset0_symmetries0_0.dat};
		\addplot[black, ultra thick, solid, opacity=0.25] table[x index=0, y index = 1] {figs/data/RES_SF9_Pi-Inf_OS1_phaseOffset0_symmetries0_0.dat};
		\addplot[black, ultra thick, solid, opacity=0.25] table[x index=0, y index = 1] {figs/data/RES_SF10_Pi-Inf_OS1_phaseOffset0_symmetries0_0.dat};
		\addplot[black, ultra thick, solid, opacity=0.25] table[x index=0, y index = 1] {figs/data/RES_SF11_Pi-Inf_OS1_phaseOffset0_symmetries0_0.dat};
		\addplot[black, ultra thick, solid, opacity=0.25] table[x index=0, y index = 1] {figs/data/RES_SF12_Pi-Inf_OS1_phaseOffset0_symmetries0_0.dat};

	\end{semilogyaxis}

\end{tikzpicture}%
  \caption{Symbol error rate of the LoRa modulation under AWGN and same-SF interference for $\text{SF}
  \in \left\{7,\hdots,12\right\}$ and $P_I = {-}3$ dB. Black dotted lines show the SER when ignoring the phase offset $\omega$ and thick transparent lines show the SER when there is only AWGN for comparison (taken from Fig.~\ref{fig:serawgn}).}
  \label{fig:serintphase}
\end{figure}
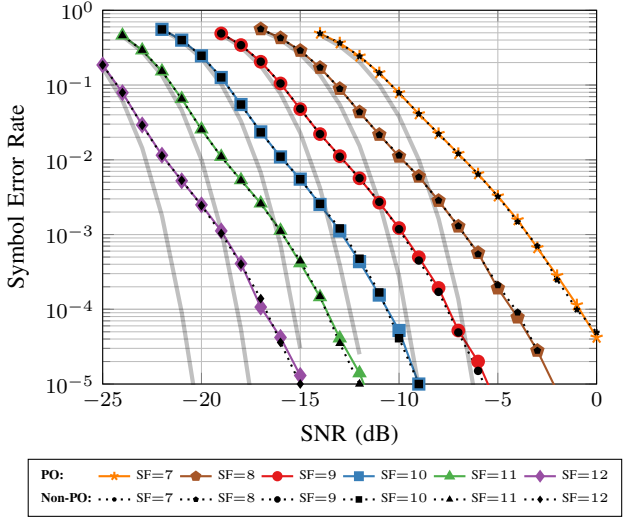

\begin{figure}[t]
  \centering
  \begin{tikzpicture}

	\small

	\begin{semilogyaxis}[
		width = \figurewidth\columnwidth,
		height = \figureheight\columnwidth,
		xlabel = {SNR (dB)},
		ylabel = {Symbol Error Rate},
		label style={font=\small},
    tick label style={font=\footnotesize},
		ylabel near ticks,
		xlabel near ticks,
		xmin = -22, xmax = -5,
		ymin = 1e-5, ymax = 1,
		grid = both,
		legend image post style={scale=0.6},
	]


		\addplot[Set1-7-1, thick, solid, mark=star, mark options={scale=1.2}] table[x index=0, y index = 1] {figs/data/RES_SF9_Pi-3.00_OS1_phaseOffset0_symmetries0_0.dat};
		\label{SF9MCPACA}
		\addplot[Set1-7-2, thick, solid, mark=pentagon*, mark options={scale=1.2}] table[x index=0, y index = 1] {figs/data/RES_SF10_Pi-3.00_OS1_phaseOffset0_symmetries0_0.dat};
		\label{SF10MCPACA}
		\addplot[Set1-7-3, thick, solid, mark=diamond*, mark options={scale=1.3}] table[x index=0, y index = 1] {figs/data/RES_SF11_Pi-3.00_OS1_phaseOffset0_symmetries0_0.dat};
		\label{SF11MCPACA}

		\addplot[black, thick, dotted, mark=star, mark options={scale=0.6, solid}] table[x index=0, y index = 1] {figs/data/APP_SF9_Pi-3.00_OS1.dat};
		\label{SF9APPACA}
		\addplot[black, thick, dotted, mark=pentagon*, mark options={scale=0.6, solid}] table[x index=0, y index = 1] {figs/data/APP_SF10_Pi-3.00_OS1.dat};
		\label{SF10APPACA}
		\addplot[black, thick, dotted, mark=diamond*, mark options={scale=0.6, solid}] table[x index=0, y index = 1] {figs/data/APP_SF11_Pi-3.00_OS1.dat};
		\label{SF11APPACA}


		\addplot[Set1-7-1, thick, solid, mark=*, mark options={scale=1.2}] table[x index=0, y index = 1] {figs/data/RES_SF9_Pi-3.00_OS10_phaseOffset0_symmetries0_Rayleigh0_0.dat};
		\label{SF9MCPNCN}
		\addplot[Set1-7-2, thick, solid, mark=square*, mark options={scale=1.05}] table[x index=0, y index = 1] {figs/data/RES_SF10_Pi-3.00_OS10_phaseOffset1_symmetries0_0.dat};
		\label{SF10MCPNCN}
		\addplot[Set1-7-3, thick, solid, mark=triangle*, mark options={scale=1.3}] table[x index=0, y index = 1] {figs/data/RES_SF11_Pi-3.00_OS10_phaseOffset1_symmetries0_0.dat};
		\label{SF11MCPNCN}

		\addplot[black, thick, dotted, mark=*, mark options={scale=0.6, solid}] table[x index=0, y index = 1] {figs/data/APP_SF9_Pi-3.00_OS3.dat};
		\label{SF9APPNCN}
		\addplot[black, thick, dotted, mark=square*, mark options={scale=0.6, solid}] table[x index=0, y index = 1] {figs/data/APP_SF10_Pi-3.00_OS3.dat};
		\label{SF10APPNCN}
		\addplot[black, thick, dotted, mark=triangle*, mark options={scale=0.6, solid}] table[x index=0, y index = 1] {figs/data/APP_SF11_Pi-3.00_OS3.dat};
		\label{SF11APPNCN}

		\node [draw,fill=white,inner sep=2pt] at (rel axis cs: 0.2,0.15) {
		\tiny
		\begin{tabular}{lcc}
			\multicolumn{3}{c}{\scriptsize\underline{Non-aligned {\tiny(this work)}}} \\
								& MC 								& Approx. \\
			SF${=}9$:	 	& \ref{SF9MCPNCN} 	&	\ref{SF9APPNCN} \\
			SF${=}10$: 	& \ref{SF10MCPNCN} & \ref{SF10APPNCN} \\
			SF${=}11$: 	& \ref{SF11MCPNCN} & \ref{SF11APPNCN}
		\end{tabular}};

		\node [draw,fill=white,inner sep=2pt] at (rel axis cs: 0.8,0.85) {
		\tiny
		\begin{tabular}{lcc}
			\multicolumn{3}{c}{\scriptsize\underline{Aligned~{\tiny\cite{Elshabrawy2018b}}} } \\
								& MC 								& Approx. \\
			SF${=}9$:	 	& \ref{SF9MCPACA} 	&	\ref{SF9APPACA} \\
			SF${=}10$: 	& \ref{SF10MCPACA} & \ref{SF10APPACA} \\
			SF${=}11$: 	& \ref{SF11MCPACA} & \ref{SF11APPACA}
		\end{tabular}};

	\end{semilogyaxis}

\end{tikzpicture}%
  \caption{Symbol error rate of the LoRa modulation under AWGN and same-SF interference for $\text{SF}
\in \left\{9,10,11\right\}$ and $P_I = {-}3$ dB. The approximations of~\cite{Elshabrawy2018b} and \eqref{eq:approxfinal} are shown with black dotted lines.}
  \label{fig:serint}
\end{figure}
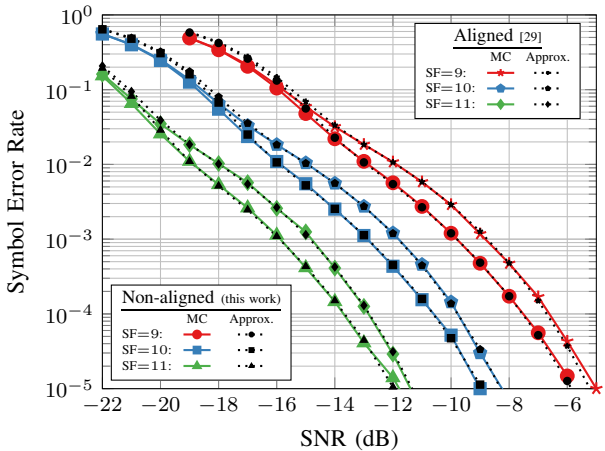

\begin{figure}[t]
  \centering
  \begin{tikzpicture}

	\small

	\begin{axis}[
		width = \figurewidth\columnwidth,
		height = \figureheight\columnwidth,
		xlabel = {SIR (dB)},
		ylabel = {SNR (dB)},
		label style={font=\small},
    tick label style={font=\footnotesize},
		xtick distance=1,
		ytick distance=5,
		ylabel near ticks,
		xlabel near ticks,
		xmin = 0, xmax = 10,
		ymin = -22, ymax = 16,
		grid = both,
		legend image post style={scale=0.6},
	]

		\addplot[Set1-7-5, thick, dashed, mark=star, mark options={scale=1.2}] table[x index=0, y index = 1] {figs/data/APP_IntTau_SNR_SIR_SF7_Pi-0.50_OS1.dat};
		\label{SF7APPACA}
		\addplot[Set1-7-4, thick, dashed, mark=pentagon*, mark options={scale=1.2}] table[x index=0, y index = 1] {figs/data/APP_IntTau_SNR_SIR_SF12_Pi-0.50_OS1.dat};
		\label{SF12APPACA}

		\addplot[Set1-7-5, thick, solid, mark=*, mark options={scale=1.2, solid}] table[x index=0, y index = 1] {figs/data/APP_NonIntTau_SNR_SIR_SF7_Pi-0.50_OS5.dat};
		\label{SF7APPNCN}
		\addplot[Set1-7-4, thick, solid, mark=square*, mark options={scale=1.05, solid}] table[x index=0, y index = 1] {figs/data/APP_NonIntTau_SNR_SIR_SF12_Pi-0.50_OS4.dat};
		\label{SF12APPNCN}

		\node [draw,fill=white,inner sep=2pt] at (rel axis cs: 0.42,0.85) {
		\scriptsize
		\begin{tabular}{lcc}
			\multicolumn{2}{c}{\footnotesize\underline{Non-aligned {\scriptsize(this work)}}} \\
						 & Approx. \\
			SF${=}7$:	 	 &	\ref{SF7APPNCN} \\
			SF${=}12$: 	 &  \ref{SF12APPNCN}
		\end{tabular}};

		\node [draw,fill=white,inner sep=2pt] at (rel axis cs: 0.82,0.85) {
		\scriptsize
		\begin{tabular}{lcc}
			\multicolumn{2}{c}{\footnotesize\underline{Aligned~{\scriptsize\cite{Elshabrawy2018b}}} } \\
						& Approx. \\
			SF${=}7$:	 	&	\ref{SF7APPACA} \\
			SF${=}12$: 	&   \ref{SF12APPACA}
		\end{tabular}};

	\end{axis}

\end{tikzpicture}%
  \caption{Required SNR for a target symbol error rate of $2\cdot 10^{-5}$ as a function of the SIR for $\text{SF}=7$ and $\text{SF}=12$.}
  \label{fig:SNR_SIR}
\end{figure}
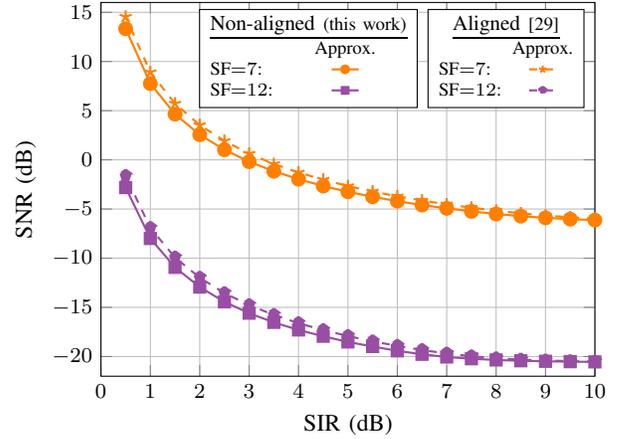

In Fig.~\ref{fig:serint}, we show the results of a Monte Carlo simulation for the SER of a LoRa user with $\text{SF} \in \left\{9,10,11\right\}$ using the chip-aligned model of~\cite{Elshabrawy2018b} and the  model we described in this work, as well as the corresponding approximations in~\eqref{eq:approxint} and~\cite{Elshabrawy2018b}, respectively. We observe that there exists a significant difference of approximately $1$~dB between the two models, meaning that the chip-aligned model of ~\cite{Elshabrawy2018b} is pessimistic in the computation of the SER. This can be intuitively explained as follows. When the offset $\tau$ is an integer, the maximum value of the interference magnitudes $A_{k,1}$ and $A_{k,2}$ with respect to the index $k$ is always larger than when $\tau$ is not an integer. As such, considering only chip-aligned interference is a worst-case scenario. Finally, we can clearly observe that the low-complexity computation of~\eqref{eq:ser_full} using the approximation derived in Section~\ref{sec:SER_approx} is very accurate.

In Fig.~\ref{fig:SNR_SIR}, we show the required SNR for a target SER performance of $2\cdot 10^{-5}$, for different SIR levels and for the two extremal spreading factors, i.e., $\text{SF} = 7$ and $\text{SF} = 12$. The target SER performance of $2\cdot 10^{-5}$ is chosen similarly to~\cite{Elshabrawy2018b} and in accordance with the sensitivity thresholds for a bandwidth of $ B=125$~kHz and a coding rate of $R=\nicefrac{4}{5} $, as provided by Semtech~\cite{SX127x}. The SNR vs SIR plot is important for every framework that considers AWGN and interference jointly and was introduced in~\cite{Elshabrawy2018b}. It is obvious that as the interference power increases, there is a significant increase in the required SNR to obtain the same performance. We can observe that the chip-aligned model of~\cite{Elshabrawy2018b} overestimates the required SNR increase for both the extremal spreading factors. Moreover, the overestimation is slightly more pronounced for higher levels of interference, since in the interference-limited regime, the impact of the overestimation of the chip-aligned model is higher than in the noise-limited regime.

It is important to note that the framework that treats the signal and the interference in a unified fashion, which is used both in~\cite{Elshabrawy2018b, Elshabrawy2019} and in this work, does not contradict, but rather enhances the view of the non-unified framework of~\cite{Georgiou2017,Bor2016, Goursaud2015}. In particular, in the non-unified framework, the SNR and the SIR are considered independently, and a LoRa message is received successfully only if both of the following two assumptions hold
\begin{enumerate}
\item $ \text{SNR} > \text{SNR}_{\text{thr}}^{(\text{SF})}$, where $\text{SNR}_{\text{thr}}^{(\text{SF})}$ is the SF-specific SNR threshold for a given target error probability, and
\item $ \text{SIR} > 6 $~dB
\end{enumerate}
In the unified SINR framework, a message can potentially survive even if $ \text{SIR} < 6 $~dB if in turn the SNR is high enough. This SNR-SIR trade-off is in essence the information that Fig.~\ref{fig:SNR_SIR} provides. Therefore, the unified SINR framework allows for a softer decision threshold on the successful reception of a LoRa message, rather than the hard 6~dB SIR threshold that is commonly used in the literature. On the other hand, the effect of multiple same-SF interferers can easily be handled under the non-unified framework. As explained in~\cite{Georgiou2017}, the analysis for multiple same-SF interferers can be simplified by considering only the most powerful interferer in the SIR value; the larger the number of same-SF users in the network, the higher the probability that the most powerful interferer will lead to a critical SIR (e.g., $ < 6 $~dB). The effect of multiple same-SF interferers on the error rate under the unified SINR framework is a field that needs to be explored further.

\subsection{Frame Error Rate}
\begin{figure}[t]
  \centering
  \begin{tikzpicture}

	\small

	\begin{semilogyaxis}[
		width = \figurewidth\columnwidth,
		height = \figureheight\columnwidth,
		xlabel = {SNR (dB)},
		ylabel = {Frame Error Rate},
		label style={font=\small},
    tick label style={font=\footnotesize},
		ylabel near ticks,
		xlabel near ticks,
		xmin = -22, xmax = -5,
		ymin = 1e-4, ymax = 1,
		grid = both,
		legend image post style={scale=0.6},
	]


		\addplot[Set1-7-1, thick, solid, mark=star, mark options={scale=1.2}] table[x index=0, y index = 1] {figs/data/FER_RES_SF9_packetLen10_Pi-3.00_OS1_phaseOffset0_0.dat};
		\label{SF9MCPACAFER}
		\addplot[Set1-7-2, thick, solid, mark=pentagon*, mark options={scale=1.2}] table[x index=0, y index = 1] {figs/data/FER_RES_SF10_packetLen10_Pi-3.00_OS1_phaseOffset0_0.dat};
		\label{SF10MCPACAFER}
		\addplot[Set1-7-3, thick, solid, mark=diamond*, mark options={scale=1.3}] table[x index=0, y index = 1] {figs/data/FER_RES_SF11_packetLen10_Pi-3.00_OS1_phaseOffset0_0.dat};
		\label{SF11MCPACAFER}

		\addplot[black, thick, dotted, mark=star, mark options={scale=0.6, solid}] table[x index=0, y index = 1] {figs/data/APP_PERslide_ours_SF9_FrLen10_Pi-3.00_OS1.dat};
		\label{SF9APPACAFER}
		\addplot[black, thick, dotted, mark=pentagon*, mark options={scale=0.6, solid}] table[x index=0, y index = 1] {figs/data/APP_PERslide_ours_SF10_FrLen10_Pi-3.00_OS1.dat};
		\label{SF10APPACAFER}
		\addplot[black, thick, dotted, mark=diamond*, mark options={scale=0.6, solid}] table[x index=0, y index = 1] {figs/data/APP_PERslide_ours_SF11_FrLen10_Pi-3.00_OS1.dat};
		\label{SF11APPACAFER}


		\addplot[Set1-7-1, thick, solid, mark=*, mark options={scale=1.2}] table[x index=0, y index = 1] {figs/data/FER_RES_SF9_packetLen10_Pi-3.00_OS10_phaseOffset1_Fri_06.09.19_0.dat};
		\label{SF9MCPNCNFER}
		\addplot[Set1-7-2, thick, solid, mark=square*, mark options={scale=1.05}] table[x index=0, y index = 1] {figs/data/FER_RES_SF10_packetLen10_Pi-3.00_OS10_phaseOffset0_corrected_0.dat};
		\label{SF10MCPNCNFER}
		\addplot[Set1-7-3, thick, solid, mark=triangle*, mark options={scale=1.3}] table[x index=0, y index = 1] {figs/data/FER_RES_SF11_packetLen10_Pi-3.00_OS10_phaseOffset0_0.dat};
		\label{SF11MCPNCNFER}

		\addplot[black, thick, dotted, mark=*, mark options={scale=0.6, solid}] table[x index=0, y index = 1] {figs/data/APP_PERslide_ours_SF9_FrLen10_Pi-3.00_OS5.dat};
		\label{SF9APPNCNFER}
		\addplot[black, thick, dotted, mark=square*, mark options={scale=0.6, solid}] table[x index=0, y index = 1] {figs/data/APP_PERslide_ours_SF10_FrLen10_Pi-3.00_OS5.dat};
		\label{SF10APPNCNFER}
		\addplot[black, thick, dotted, mark=triangle*, mark options={scale=0.6, solid}] table[x index=0, y index = 1] {figs/data/APP_PERslide_ours_SF11_FrLen10_Pi-3.00_OS5.dat};
		\label{SF11APPNCNFER}

		\node [draw,fill=white,inner sep=2pt] at (rel axis cs: 0.2,0.15) {
		\tiny
		\begin{tabular}{lcc}
			\multicolumn{3}{c}{\scriptsize\underline{Non-aligned {\tiny(this work)}}} \\
								& MC 								& Approx. \\
			SF${=}9$:	 	& \ref{SF9MCPNCNFER} 	&	\ref{SF9APPNCNFER} \\
			SF${=}10$: 	& \ref{SF10MCPNCNFER} & \ref{SF10APPNCNFER} \\
			SF${=}11$: 	& \ref{SF11MCPNCNFER} & \ref{SF11APPNCNFER}
		\end{tabular}};

		\node [draw,fill=white,inner sep=2pt] at (rel axis cs: 0.8,0.85) {
		\tiny
		\begin{tabular}{lcc}
			\multicolumn{3}{c}{\scriptsize\underline{Aligned~{\tiny\cite{Elshabrawy2018b}}} } \\
								& MC 								& Approx. \\
			SF${=}9$:	 	& \ref{SF9MCPACAFER} 	&	\ref{SF9APPACAFER} \\
			SF${=}10$: 	& \ref{SF10MCPACAFER} & \ref{SF10APPACAFER} \\
			SF${=}11$: 	& \ref{SF11MCPACAFER} & \ref{SF11APPACAFER}
		\end{tabular}};

	\end{semilogyaxis}

\end{tikzpicture}%
  \caption{Frame error rate of the LoRa modulation for a frame length $F=10$ under AWGN and same-SF interference for $\text{SF}
\in \left\{9,10,11\right\}$ and $P_I = {-}3$ dB. The approximations of~\cite{Elshabrawy2018b} and \eqref{eq:approxfinal} are shown with black dotted lines.}
  \label{fig:FERIntSliding}
\end{figure}
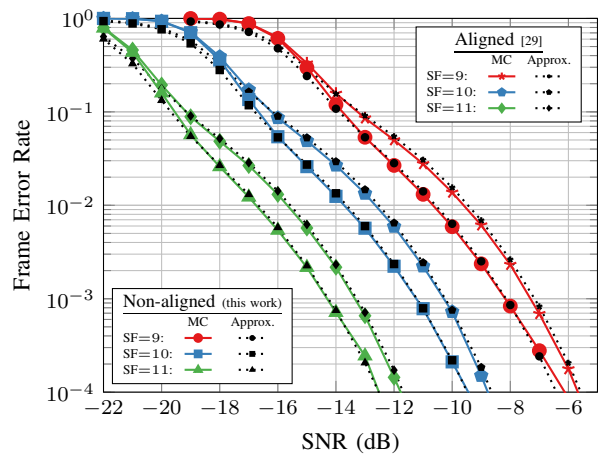

In Fig.~\ref{fig:FERIntSliding}, we show the results of a Monte Carlo simulation for the FER of a LoRa user with $\text{SF} \in \left\{9,10,11\right\}$ using the chip-aligned model of~\cite{Elshabrawy2018b} and the model we described in this work, as well as the corresponding approximation described in Section~\ref{sec:fer}. The frame contains $F=10$ LoRa symbols, which is a valid data payload length for LoRa. We observe the same difference of approximately $1$~dB between the two models. Moreover, we can see that the approximation for the FER described in Section~\ref{sec:fer} is very accurate.

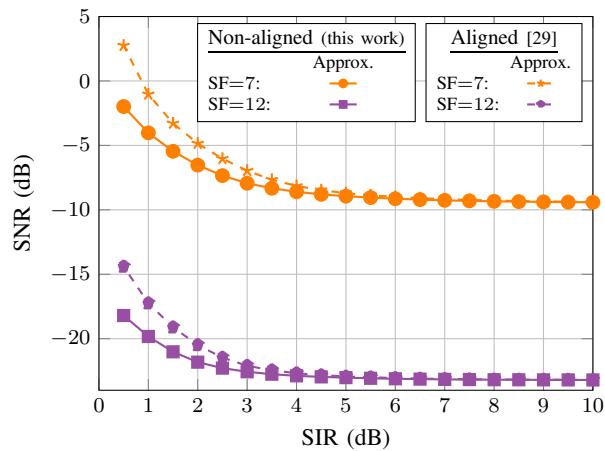
\begin{figure}[t]
  \centering
  \begin{tikzpicture}

	\small

	\begin{axis}[
		width = \figurewidth\columnwidth,
		height = \figureheight\columnwidth,
		xlabel = {SIR (dB)},
		ylabel = {SNR (dB)},
		label style={font=\small},
    tick label style={font=\footnotesize},
		xtick distance=1,
		ytick distance=5,
		ylabel near ticks,
		xlabel near ticks,
		xmin = 0, xmax = 10,
		ymin = -24, ymax = 5,
		grid = both,
		legend image post style={scale=0.6},
	]

		\addplot[Set1-7-5, thick, dashed, mark=star, mark options={scale=1.2}] table[x index=0, y index = 1] {figs/data/APP_slide_Elsh_SNR_SIR_SF7_FrLen10_Pi-0.50_OS1_PERtarget0.10.dat};
		\label{SF7APPACA}
		\addplot[Set1-7-4, thick, dashed, mark=pentagon*, mark options={scale=1.2}] table[x index=0, y index = 1] {figs/data/APP_slide_Elsh_SNR_SIR_SF12_FrLen10_Pi-0.50_OS1_PERtarget0.10.dat};
		\label{SF12APPACA}

		\addplot[Set1-7-5, thick, solid, mark=*, mark options={scale=1.2, solid}] table[x index=0, y index = 1] {figs/data/APP_slide_ours_SNR_SIR_SF7_FrLen10_Pi-0.50_OS5_PERtarget0.10.dat};
		\label{SF7APPNCN}
		\addplot[Set1-7-4, thick, solid, mark=square*, mark options={scale=1.05, solid}] table[x index=0, y index = 1] {figs/data/APP_slide_ours_SNR_SIR_SF12_FrLen10_Pi-0.50_OS5_PERtarget0.10.dat};
		\label{SF12APPNCN}

		\node [draw,fill=white,inner sep=2pt] at (rel axis cs: 0.42,0.85) {
		\scriptsize
		\begin{tabular}{lcc}
			\multicolumn{2}{c}{\footnotesize\underline{Non-aligned {\scriptsize(this work)}}} \\
						 & Approx. \\
			SF${=}7$:	 	 &	\ref{SF7APPNCN} \\
			SF${=}12$: 	 &  \ref{SF12APPNCN}
		\end{tabular}};

		\node [draw,fill=white,inner sep=2pt] at (rel axis cs: 0.82,0.85) {
		\scriptsize
		\begin{tabular}{lcc}
			\multicolumn{2}{c}{\footnotesize\underline{Aligned~{\scriptsize\cite{Elshabrawy2018b}}} } \\
						& Approx. \\
			SF${=}7$:	 	&	\ref{SF7APPACA} \\
			SF${=}12$: 	&   \ref{SF12APPACA}
		\end{tabular}};

	\end{axis}

\end{tikzpicture}%
  \caption{Required SNR for a target frame error rate of $10^{-1}$ as a function of the SIR for $\text{SF}=7$ and $\text{SF}=12$.}
  \label{fig:SNR_SIR_FERtarget}
\end{figure}

In Fig.~\ref{fig:SNR_SIR_FERtarget}, we show the required SNR for a target FER performance of $10^{-1}$, for different SIR levels, and for the two extremal spreading factors, i.e., $\text{SF} = 7$ and $\text{SF} = 12$. Similarly to Fig.~\ref{fig:SNR_SIR}, as the interference power increases, there is an increase in the required SNR to obtain the same performance. For the chosen target FER performance of $10^{-1}$ we can observe that the overestimation of the chip-aligned model of~\cite{Elshabrawy2018b} is clearly more pronounced for higher levels of interference.

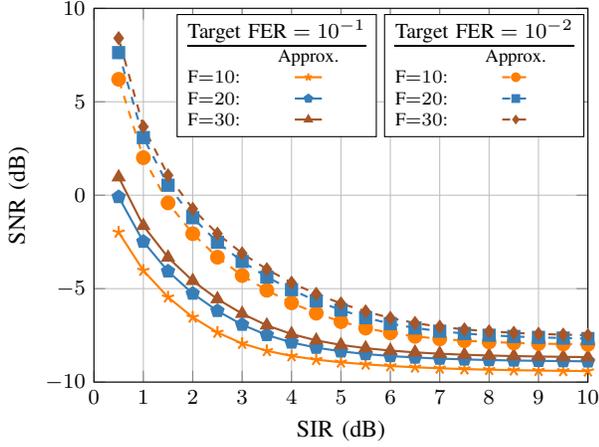
\begin{figure}[t]
  \centering
  \begin{tikzpicture}

	\small

	\begin{axis}[
		width = \figurewidth\columnwidth,
		height = \figureheight\columnwidth,
		xlabel = {SIR (dB)},
		ylabel = {SNR (dB)},
		label style={font=\small},
    tick label style={font=\footnotesize},
		xtick distance=1,
		ytick distance=5,
		ylabel near ticks,
		xlabel near ticks,
		xmin = 0, xmax = 10,
		ymin = -10, ymax = 10,
		grid = both,
		legend image post style={scale=0.6},
	]

		\addplot[Set1-7-5, thick, solid, mark=star, mark options={scale=1.2}] table[x index=0, y index = 1] {figs/data/APP_slide_ours_SNR_SIR_SF7_FrLen10_Pi-0.50_OS5_PERtarget0.10.dat};
		\label{F10APPhighFER}
		\addplot[Set1-7-2, thick, solid, mark=pentagon*, mark options={scale=1.2}] table[x index=0, y index = 1] {figs/data/APP_slide_ours_SNR_SIR_SF7_FrLen20_Pi-0.50_OS5_PERtarget0.10.dat};
		\label{F20APPhighFER}
		\addplot[Set1-7-6, thick, solid, mark=triangle*, mark options={scale=1.2}] table[x index=0, y index = 1] {figs/data/APP_slide_ours_SNR_SIR_SF7_FrLen30_Pi-0.50_OS5_PERtarget0.10.dat};
		\label{F30APPhighFER}

		\addplot[Set1-7-5, thick, dashed, mark=*, mark options={scale=1.2, solid}] table[x index=0, y index = 1] {figs/data/APP_slide_ours_SNR_SIR_SF7_FrLen10_Pi-0.50_OS5_PERtarget0.01.dat};
		\label{F10APPlowFER}
		\addplot[Set1-7-2, thick, dashed, mark=square*, mark options={scale=1.05, solid}] table[x index=0, y index = 1] {figs/data/APP_slide_ours_SNR_SIR_SF7_FrLen20_Pi-0.50_OS5_PERtarget0.01.dat};
		\label{F20APPlowFER}
		\addplot[Set1-7-6, thick, dashed, mark=diamond*, mark options={scale=1.05, solid}] table[x index=0, y index = 1] {figs/data/APP_slide_ours_SNR_SIR_SF7_FrLen30_Pi-0.50_OS5_PERtarget0.01.dat};
		\label{F30APPlowFER}

		\node [draw,fill=white,inner sep=2pt] at (rel axis cs: 0.37,0.82) {
		\scriptsize
		\begin{tabular}{lcc}
			\multicolumn{2}{c}{\footnotesize\underline{Target $\text{FER}=10^{-1}$}} \\
						 & Approx. \\
			F${=}10$:	 	 &	\ref{F10APPhighFER} \\
			F${=}20$: 	 &  \ref{F20APPhighFER} \\
			F${=}30$: 	 &  \ref{F30APPhighFER}
		\end{tabular}};

		\node [draw,fill=white,inner sep=2pt] at (rel axis cs: 0.79,0.82) {
		\scriptsize
		\begin{tabular}{lcc}
			\multicolumn{2}{c}{\footnotesize\underline{Target $\text{FER}=10^{-2}$}} \\
						 & Approx. \\
			F${=}10$:	 	 &	\ref{F10APPlowFER} \\
			F${=}20$: 	 &  \ref{F20APPlowFER} \\
			F${=}30$: 	 &  \ref{F30APPlowFER}
		\end{tabular}};

	\end{axis}

\end{tikzpicture}%
  \caption{Required SNR for target frame error rates of $10^{-1}$ and $10^{-2}$, for frame lengths $F = 10, 20, 30$, as a function of the SIR for $\text{SF}=7$.}
  \label{fig:SNR_SIR_F10_20_30}
\end{figure}

Finally, in Fig.~\ref{fig:SNR_SIR_F10_20_30}, we show the required SNR for two different target FERs and three different frame lengths, for different SIR levels and for $\text{SF} = 7$. Specifically, we choose both a typical target FER performance of $10^{-1}$ and a stricter target FER of $10^{-2}$~\cite{Croce2018}, and we choose frames of length $F=10$, $F=20$, and $F=30$ LoRa symbols. We observe that longer frames require a larger increase in the required SNR for successful reception under the same interference power. Moreover, the performance requirement plays an important role, since the increase in the required SNR values for a packet to survive under a stricter target FER performance is very pronounced.

\section{Conclusion} \label{sec:conclusion}
In this work, we introduced a LoRa interference model where the interference is neither chip- nor phase-aligned with the LoRa signal-of-interest and we derived a corresponding expression for the SER and the FER. Moreover, we proved two properties of same-SF LoRa-induced interference that enabled us to reduce the complexity of calculating the SER and the FER by a factor that is approximately equal to the LoRa symbol length $N$. Finally, we derived a low-complexity approximation for both the SER and the FER and we showed that ignoring the non-integer time offsets overestimates the error rate by $1$~dB.

\appendix[]

\begin{IEEEproof}[Proof of Proposition \ref{prop:Prop_delta}]
We will first show that $\mathcal{Y}_{I_1}$ indeed contains $N{-}(s_{I_1}{-}s_{I_2})$ equivalent interference patterns for a given $s_{I_1}$ and $s_{I_2}$, which can be obtained by setting $s'_{I_1} = [s_{I_1}+\delta]_N$, $s'_{I_2} = [s_{I_2}+\delta]_N$, and $k' = [k+\delta]_N$. Recall that for $|R_{k'}|$ we have
\begin{align}
  |R_{k'}| & = \sqrt{(A'_{k,1})^{2} + (A'_{k,2})^{2} + 2A'_{k,1}A'_{k,2}\cos(\theta'_{k,1} - \theta'_{k,2})}.
\end{align}
Then, our goal essentially is to show that $|R_{k'}| = |R_{k}|$ for all $\delta$ such that $s'_{I_1} \geq s'_{I_2}$, and for all $k$.

For a given $s_{I_1}$ and $s_{I_2}$, the condition $s'_{I_1} \geq s'_{I_2}$ holds in the following two cases
\begin{align}
  \text{Case A: }  & s_{I_1} + \delta < N \text{ and } s_{I_2} + \delta < N, \\
  \text{Case B: }  & s_{I_1} + \delta \geq N \text{ and } s_{I_2} + \delta \geq N.
\end{align}
As such, it is straightforward to see that $\left|\mathcal{Y}_{I_1}\right| = N-(s_{I_1}-s_{I_2})$, meaning that~\eqref{eq:card1} holds.

We show the remainder of the proof only for case A, but it can easily be extended to case B using the same arguments. For $A_{k,i}'$, we have
\begin{align}
A_{k,i}'  & = \frac{\sin \left( \frac{\pi}{N} ([s_{I_{i}}{+}\delta]_{N}-[k{+}\delta]_{N}-\tau)\lceil \tau \rceil \right)}{\sin \left( \frac{\pi}{N} ([s_{I_{i}}{+}\delta]_{N}-[k{+}\delta]_{N}-\tau) \right)} \\
       & = \begin{cases}
         \frac{\sin \left( \frac{\pi}{N} (s_{I_{i}}-k-\tau)\lceil \tau \rceil \right)}{\sin \left( \frac{\pi}{N} (s_{I_{i}}-k-\tau) \right)}, & k+\delta < N,\\
         \frac{\sin \left( \frac{\pi}{N} (s_{I_{i}}-k-\tau)\lceil \tau \rceil + \pi\lceil \tau \rceil\right)}{\sin \left(\frac{\pi}{N} (s_{I_{i}}-k-\tau) + \pi\right)}, & k+\delta \geq N.
       \end{cases} \label{eq:casesAi}
\end{align}
We can rewrite~\eqref{eq:casesAi} as
\begin{align}
  A_{k,i}' &=
  \begin{cases}
  +A_{k,i}, & k+\delta < N, \\
  -A_{k,i}, & k+\delta \geq N \text{ and } \lceil \tau \rceil\text{ even}, \\
  +A_{k,i}, & k+\delta \geq N \text{ and } \lceil \tau \rceil\text{ odd}.
  \end{cases}
\end{align}
This means that $ A_{k,1}'^2 =  A_{k,1}^2 $, $ A_{k,2}'^2 =  A_{k,2}^2 $, and $ A_{k,1}'A_{k,2}' =  A_{k,1}A_{k,2} $ for any $k$ and $\delta$. For $\cos(\theta_{k,1}'{-}\theta_{k,2}')$, using the assumption $s_{I_{1}} \geq s_{I_{2}}$, we have
{
  \small
  \begin{align}\label{eq:cos_prime}
    \cos(\theta_{k,1}'{-}\theta_{k,2}') & = \cos\left( \frac{\pi}{N} \left( (\lambda{-}L)N {+} (s_{I_{1}} {-} s_{I_{2}})(\tau{-}\lceil \tau \rceil{+}1) {+} \right. \right. \nonumber \\
                                & + \left. \left. N(s'_{I_{2}} {-} k') \right)\right).
  \end{align}
}%
For the term $N(s'_{I_{2}} {-} k')$, using the fact that $s_{I_{2}}+\delta < N$ we have the following two cases
\begin{align}
  N(s'_{I_{2}} {-} k')  & = N((s_{I_{2}}+\delta){-}[k+\delta]_N) \\
  & = \begin{cases}
    N(s_{I_{2}}{-}k), & k+\delta < N, \\
    N(s_{I_{2}}{-}k)-N^2, & k+\delta \geq N.
  \end{cases} \label{eq:thirdtermcases}
\end{align}
Combining~\eqref{eq:cos_prime} and \eqref{eq:thirdtermcases} and using the fact that $N$ is a power of two, we can show that
\begin{align}
  \cos(\theta_{k,1}'{-}\theta_{k,2}') & = \cos(\theta_{k,1}{-}\theta_{k,2}). \label{eq:thetacase1}
\end{align}
We have thus shown the claimed result for $\mathcal{Y}_{I_1}$.

The corresponding proof for $\mathcal{Y}_{I_2}$ is omitted for the sake of brevity, but it can be obtained using the same arguments. The main differences in this case are that we no longer have $A_{k,1}'A_{k,2}' = A_{k,1}A_{k,2}$ and $\cos(\theta_{k,1}'-\theta_{k,2}') = \cos(\theta_{k,1}-\theta_{k,2})$, but it can be shown that
\begin{align}
  A_{k,1}'A_{k,2}'\cos(\theta_{k,1}'-\theta_{k,2}') & = A_{k,1}A_{k,2}\cos(\theta_{k,1}-\theta_{k,2}-2\lambda\pi). \label{eq:cosY2}
\end{align}
As such, the interference patterns are identical with each other for all $\delta$ that lead to $s'_{I_1} < s'_{I_2}$, but they are different from the interference patterns that are obtained for the $\delta$ values that lead to $s'_{I_1} \geq s'_{I_2}$.

In the special case where $\lambda = 0$ (i.e., when $\tau$ is an integer),~\eqref{eq:cosY2} becomes
\begin{align}
  A_{k,1}'A_{k,2}'\cos(\theta_{k,1}'-\theta_{k,2}') & = A_{k,1}A_{k,2}\cos(\theta_{k,1}-\theta_{k,2}),
\end{align}
meaning that all interference patterns in both $\mathcal{Y}_{I_1}$ and $\mathcal{Y}_{I_2}$ are indeed equivalent.
\end{IEEEproof}

\begin{IEEEproof}[Proof of Proposition \ref{prop:Prop_offsets}]
Let $k' = -k-(N-1) + \left[s_{I_1}+s_{I_2}\right]_N$. In order to prove~Proposition~\ref{prop:Prop_offsets}, we will to show that
\begin{align}
  |R_{k'}| = |R_{k}|, \quad k = 0,\hdots,N-1.
\end{align}
Recall that for $|R_{k'}|$ we have
\begin{align}
  |R_{k'}| & = \sqrt{(A_{k,1}')^{2} + (A_{k,2}')^{2} + 2A_{k,1}'A_{k,2}'\cos(\theta'_{1} - \theta'_{2})}.
\end{align}
For $A'_{k,1}$ we have
{
\small
\begin{align}
  A_{k,1}'  & = \frac{\cos \left(\frac{\pi}{N}\left(s_{I_1}' {-} k' {-} \tau'\right)(\lceil \tau \rceil')\right)}{\cos \left(\frac{\pi}{N}\left(s_{I_1}' {-} k' {-} \tau' \right)\right)} \\
        & = \begin{cases}
              \frac{\cos \left(\frac{\pi}{N}\left({-}s_{I_2} {+} k {+} \tau\right)(N{-}\lceil \tau \rceil)\right)}{\cos \left(\frac{\pi}{N}\left(-s_{I_2} {+} k {+} \tau\right)\right)},  & s_{I_1}{+}s_{I_2} {<} N, \\
              \frac{\cos \left(\frac{\pi}{N}\left({-}s_{I_2} {+} k {+} \tau\right)(N{-}\lceil \tau \rceil) {+} \pi (N-\lceil\tau\rceil) \right)}{\cos \left(\frac{\pi}{N}\left(-s_{I_2} {+} k {+} \tau\right){+} \pi\right)},  & s_{I_1}{+}s_{I_2} {\geq} N,
            \end{cases} \label{eq:casesA1}
\end{align}%
}%
where $\lceil \tau \rceil' = N-\lceil\tau\rceil$. Note that $\lceil \tau \rceil' \neq \lceil \tau'\rceil$ because the term $\lceil \tau \rceil'$ comes from the cardinality of $\mathcal{N}_{L_2}$, which is $N-\lceil\tau\rceil$, and not from the time shift $\tau'$. We can rewrite~\eqref{eq:casesA1} as
\begin{align}
  A_{k,1}' &=
  \begin{cases}
  +A_{k,2}, & s_{I_1}{+}s_{I_2} {<} N, \\
  -A_{k,2}, & s_{I_1}{+}s_{I_2} {\geq} N\text{ and } \lceil \tau \rceil \text{ even}, \\
  +A_{k,2}, & s_{I_1}{+}s_{I_2} {\geq} N \text{ and } \lceil \tau \rceil \text{ odd}.
\end{cases} \label{eq:casesA1final}
\end{align}
The equivalent expression for $A_{k,2}'$ can be obtained by exchanging the subscripts $1$ and $2$ in~\eqref{eq:casesA1final}. This means that $ A_{k,1}'^2 =  A_{k,2}^2 $, $ A_{k,2}'^2 =  A_{k,1}^2 $, and $ A_{k,1}'A_{k,2}' =  A_{k,2}A_{k,1} $. Finally, for $\cos (\theta_{k,1}'-\theta_{k,2}')$, using the fact that~\eqref{eq:deltalprime} implies $L' = N-2-L$ and $\lambda' =  1-\lambda$, we have
{
\small
\begin{align}
  \cos (\theta_{k,1}'{-}\theta_{k,2}')  & = \cos \left(\frac{\pi}{N}\left( (\lambda'{-}L')N {+} \right.\right. \nonumber \\
                                & + \left.\left. (s_{I_1}'{-}s_{I_2}')(2\tau' {-} \lceil \tau\rceil'{+}1){+}(s_{I_2}'{-}k')N \right)\right) \\
                                & = \cos \left(\frac{\pi}{N}\left( ({-}\lambda{+}L)N {+} (s_{I_2} {-} [s_{I_1}{+}s_{I_2}]_N+k)N {+} \right.\right. \nonumber \\
                                & + \left.\left. (s_{I_1}{-}s_{I_2})({-}2\tau {+} \lceil \tau\rceil{-}1{+}N) \right) \right).
\end{align}
}%
By taking two cases for $[s_{I_1}{+}s_{I_2}]_N$, it is straightforward to show that in both cases
\begin{align}
  \cos (\theta_{k,1}'{-}\theta_{k,2}')  & = \cos (\theta_{k,1}{-}\theta_{k,2}).
\end{align}
As such, $|R_{k'}| = |R_{k}|$, $k = 0,\hdots,N-1,$ indeed holds.
\end{IEEEproof}

\balance
\bibliographystyle{IEEEtran}
\bibliography{IEEEabrv,refs}

\end{document}